\documentclass[aps,amsmath,amssymb,twocolumn,superscriptaddress,prb]{revtex4-2}
\usepackage{graphicx}
\usepackage{amsthm}
\usepackage{color}
\usepackage{upquote}

\newtheorem{theorem}{Theorem}
\newtheorem{corollary}{Corollary}[theorem]
\newtheorem{lemma}[theorem]{Lemma}

\theoremstyle{definition}
\newtheorem{definition}{Definition}

\renewcommand{\r}{\vec{r}}

\usepackage{stmaryrd}
\newcommand{\lightcone}{\leftrightarrowtriangle}

\begin{document}

\title{How to detect the spacetime curvature without rulers and clocks}

\author{A.~V.~Nenashev}
\thanks{On leave of absence from Rzhanov Institute of Semiconductor Physics
  and the Novosibirsk State University, Russia}
\email{nenashev\_isp@mail.ru}
\affiliation{Department of Physics and Material Sciences Center,
  Philipps-Universit\"at Marburg, D-35032 Marburg, Germany}

\author {S.~D.~Baranovskii}
\email{sergei.baranovski@physik.uni-marburg.de}
\affiliation{Department of Physics and Material Sciences Center,
  Philipps-Universit\"at Marburg, D-35032 Marburg, Germany}
\affiliation{Department f\"ur Chemie, Universit\"at zu K\"oln,
  Luxemburger Stra\ss e 116,
  50939 K\"{o}ln, Germany}
  

\date{\today}

\begin{abstract}
We demonstrate how one can distinguish a curved 4-dimensional spacetime from a flat one, when it is possible, relying only on the causality relations between events. 
It is known that it is possible only for spacetimes that are not conformally flat. We prove that if a spacetime is not conformally flat, then its non-flatness can be verified by only a few (sixteen) measurements of causal relations. Therefore the results of this paper clarify what can be said about flatness or non-flatness of the spacetime after a finite number of measurements of causal relations.
\end{abstract}

\maketitle   

\section{Introduction}
\label{sec:intro}

What is the flat space and what is the curved space? The usual answer is: the space is flat if it can be mapped (at least piecewise) onto a flat Euclidean space with preserving the metric. Otherwise, the space is curved. 

In more details, if one wants to know whether the space is flat or curved, one have to choose a coordinate system---a collection of coordinates for every point---and measure all the distances between neighboring points. These measurements provide the metric tensor $g_{ik}$ that generally depends on coordinates, i.~e. on radius-vector $\r$~\cite[\S13.2]{Wheeler_book}. Function $g_{ik}(\r)$ (a metric) contains all information about geometry of the space. Then, if the same function $g_{ik}(\r)$ can be achieved in the flat space, by choosing a proper coordinate system there, then the geometry of the space under study is also flat. And if it is impossible, then the properties of the given space differ from that of the flat one, meaning that the space is curved. 

A simplest example of the curved space is the sphere, and a customary coordinate system on it is defined by the latitude and the longitude. One cannot draw, on a flat piece of paper, a latitude-longitude net that reproduces the same net on the sphere without any distortions---and this exactly means that the sphere is not flat.

All the above is valid also for the \emph{spacetime}. It is flat (curved) if its metric can (cannot) be mapped onto the Minkowski space. To determine the metric tensor of the spacetime, one have to measure both distances and time intervals between events, and thus have to be equipped by some sorts of rulers and clocks. 
In principle, a finite number of measurement is enough, as in a ``five-point curvature detector'' considered by Synge~\cite[Chapter~XI, \S8]{Synge_book}. 
An alternative way of finding out the spacetime curvature consists in looking at neighboring word lines of freely falling test particles, and measuring how fast initially parallel world lines diverge in the course of time~\cite[Chapter~11]{Wheeler_book}. 
The ``gravity gradiometer'' for measuring the curvature of spacetime~\cite[\S16.5]{Wheeler_book} is a construction made of rigid rods. 
Detection of gravitational waves, i.~e. curvature variations, is based on interferometric measurements of distances between stationary objects in now-operating observatories such as LIGO~\cite{Aasi2015LIGO}, Virgo~\cite{Acernese2015Virgo} and KAGRA~\cite{Akutsu2019KAGRA}, or between freely moving objects (satellites) in proposed LISA detector~\cite{Amaro-Seoane2017LISA}. Also laser ranging of the Moon and artificial satellites, as well as binary pulsar timing, may be used for observation of low-frequency gravitational waves~\cite{Blas2022PRL}. All these methods also imply usage of rulers and/or clocks.
Even the ``ideal rods and clocks built form geodesic world lines'' (the Marzke-Wheeler clocks)~\cite[\S16.4]{Wheeler_book} are based on the method of Schild's ladder, which demands finding midpoints of geodesic line segments, i.~e. using rods or clocks. 
Quantum mechanics allows to measure the difference between proper times along two world lines in interferometric experiments with massive bodies, which leads to so-called gravitational Aharonov-Bohm effect~\cite{Hohensee2012}. Recently, this effect was observed in an experiment with freely falling rubidium atoms~\cite{Overstreet2022}. This experiment can be considered as a probe of space-time curvature, where atoms act as quantum clocks~\cite{Roura2022}.

But what if we do not have any rulers and clocks? This question arises from the fact that the spacetime, by itself, does not contain any built-in length or time standards (apart from Planck units that are too small for having any relation to ``normal'', non-quantum-gravity physics). Any information about lengths and times may come only from material objects, such as rigid rods, oscillating balance wheels or quartz crystals, and so on. 

It is even not evident, whether there are appropriate rulers (clocks) at any spatial (temporal) scale. According to the ``desert'' hypothesis in particle physics, there are no particles with masses from $\simeq 10^{12}$~eV to $10^{25}$~eV. Then, it might be no physical objects with lengths between $\simeq 10^{-18}$~m (the currently probed length scale) and $10^{-31}$~m (the Grand Unified Theory length scale). In such a case, what would be the meaning of spacetime curvature within this range of scales?

The spacetime itself, however, possesses very fundamental relations between its points (events)---namely the \emph{causal relations}. 
Roughly speaking, causal relations between two events $A$ and $B$ reflect where is event $B$ located with respect to the \emph{light cone} of event $A$. Event $B$ is located on the light cone of event $A$, if one can send a light signal at one of these events (the earliest one), and the signal will arrive exactly at the other event. 
If it is not the case, then event $B$ is located either in the \emph{absolute future} with respect to $A$ (i.~e. information from $A$ can reach $B$), of in the \emph{absolute past} (i.~e. information from $B$ can reach $A$), or \emph{elsewhere} (i.~e. it is impossible to send information neither from $A$ to $B$, nor from $B$ to $A$). All these concepts---light cone, absolute future/past, elsewhere---are causal relations, and they do not rely on any measuring tools like rulers and clocks. It is worth noting that the term ``light cone'' does not actually refer to light as a concrete physical phenomenon---an electromagnetic wave. The word ``light'' here just means a fastest possible carrier of information, irrespective of its physical nature.

\begin{figure}
\includegraphics[width=\linewidth]{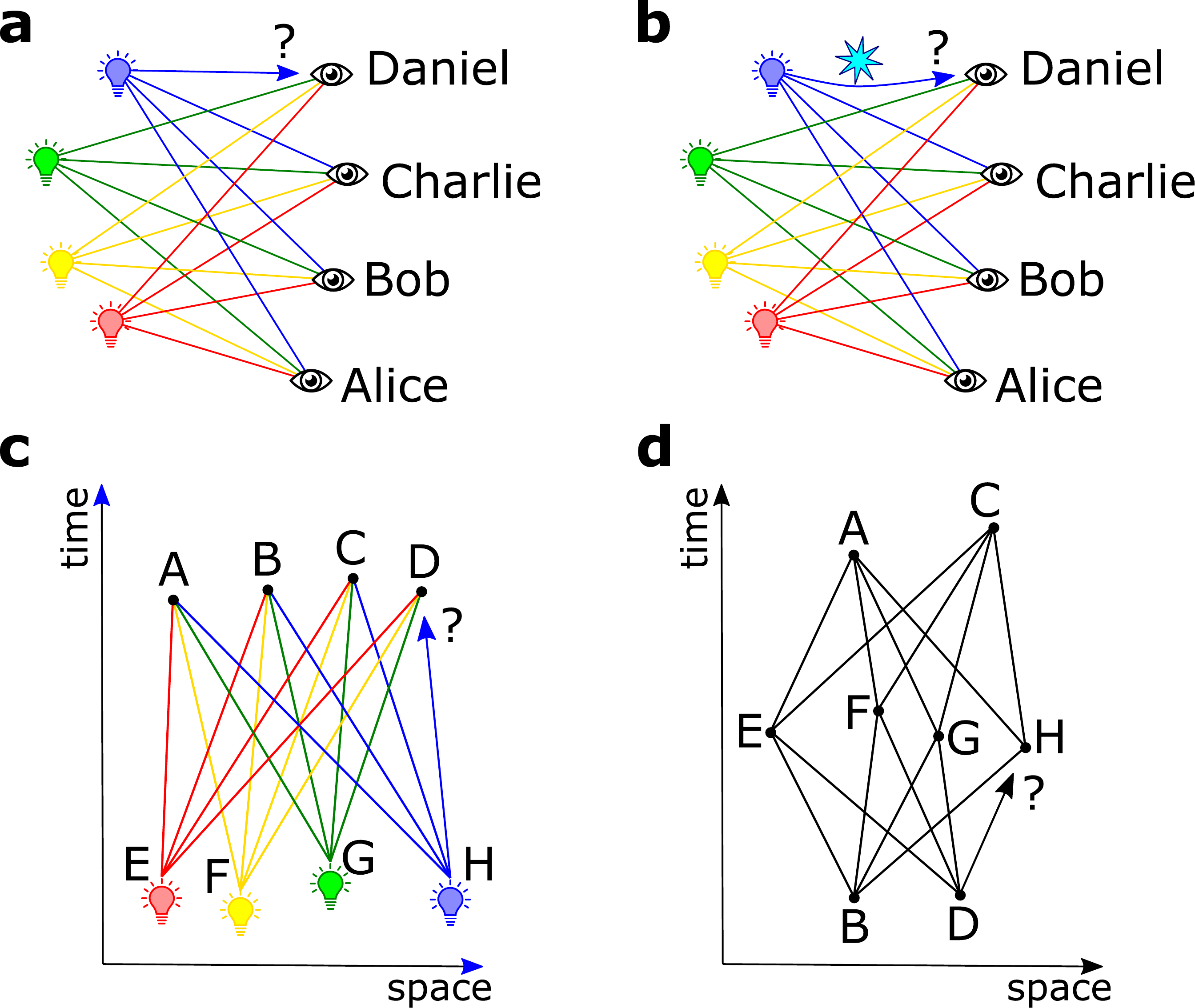}
\caption{A thought experiment with light signals: (a) in the flat Minkowski space; (b) in the space with a massive body (a star) near one of light rays; (c) a sketch of world lines of light signals; (d) a similar sketch for an alternative version of the experiment.}
\label{fig:gedanken}
\end{figure}

Let us consider a gedanken experiment, in which only causal relations between events are involved. There are four lamps: red, yellow, green and blue, and four observers: Alice, Bob, Charlie and Daniel, see Fig.~\ref{fig:gedanken}a. Each lamp was initially switched off, and switches on at some moment (these moments can be different for different lamps). Imagine that Alice saw the light from all four lamps \emph{simultaneously}. Also Bob saw all four switchings on simultaneously. And so did Charlie. It is also known that Daniel saw simultaneously the light from red, yellow and green lamps. Then, the question arises: \emph{did Daniel see the blue light at the same moment of time as he saw other three lights?}

In the flat Minkowski spacetime, where light travels with a constant velocity $c$, the answer to this question is ``yes'' (provided that all lamps and observers are located in different places). This answer follows from Theorem~\ref{th:Minkowski} that is formulated and proved below. In the terminology introduced in Section~\ref{sec:theorems}, answer ``yes'' is guarenteed by \emph{well-stitchedness} of the Minkowski spacetime.

Let us repeat this thought experiment in the curved spacetime of general relativity. It is possible that a light ray from the blue lamp to Daniel passed close to a massive body---a star, see Fig.~\ref{fig:gedanken}b. Then the ray underwent gravitational deflection of light, and arrived to Daniel later then in the case when the star is absent (so-called Shapiro time delay effect~\cite[\S40.4]{Wheeler_book}). In this circumstance, Daniel did not see all four lamps simultaneously, and the answer to the above-formulated question is ``no''.

Hence, {\bf it is possible to figure out that the spacetime is curved by testing causal relations only, without any clocks and rulers.} Namely, if the experiment considered above gave the result ``no'' at least once, then the existence of spacetime curvature is proven. This is one of the principal results of the present article.

But what if this experiment, being repeated for all possible positions of lamps and observers and all moments of switching on, always gives the result ``yes''? Does it mean that the spacetime is flat? No, it does not. 
It is well known that causal relations carry some fraction of information about the spacetime metric, but not the full information. Light cones are conserved under a \emph{conformal transformation} that acts on the metric tensor $g_{ik}$ as follows:
\begin{equation}
\label{eq:conformal}
g_{ik}(\r) \to \Omega^2(\r) g_{ik}(\r),
\end{equation}
where $\Omega$ is an arbitrary non-zero function of the spacetime coordinates~\cite[Sec.~6.13]{dInverno_book}. Therefore causal relations also remain unchanged under transformation~(\ref{eq:conformal}).  This means that one cannot distinguish between two metrics, that turn into each other by conformal transformations, on the basis of causal relations only. In particular, let us consider a \emph{conformally flat} spacetime, i.~e. that can be mapped to flat one by a conformal transformation. An example of such a conformally flat spacetime is a manifold of constant curvature. It clearly follows from the above, that {\bf a conformally flat spacetime is indistinguishable from a flat one by examining causal relations.}  

And what if the spacetime under study is not conformally flat? 
This question can be answered on the basis of Theorem~\ref{th:Weyl} that is formulated and proved in the present article. 
It follows from Theorem~\ref{th:Weyl} that {\bf each non-conformally-flat 4-dimensional spacetime can be distinguished from a flat (Minkowski) one by testing a finite set of causal relations.} This set resembles very closely that of the thought experiment considered above; it includes 16 relations between 8 events.

To summarize: the results of this paper clarify what can be said about flatness or non-flatness of the spacetime after a finite number of measurements of causal relations between events. If a 4-dimensional spacetime is not conformally flat, then its non-flatness can be verified by as few as 16 measurements of causal relations. And if a spacetime is conformally flat, then, as is known, it cannot be distinguished from a flat spacetime by any number of such measurements.

In this paper we focus on the usual, four-dimensional spacetime that possesses three spatial and one temporal coordinates.

The rest of this paper is organized as follows. 
In Section~\ref{sec:theorems}, we introduce the concept of ``well-stitched'' spacetime, defined solely in terms of causal relations. We formulate Theorems~\ref{th:Minkowski} and~\ref{th:Weyl}  that constitute the main result of this article. 
In Section~\ref{sec:discussion}, we discuss various issues regarding to the results of this paper, in particular, possibility of practical detection of the spacetime curvature using the proposed scheme. 
In Section~\ref{sec:proof1}, we prove Theorem~\ref{th:Minkowski} that applies to light cones in the flat Minkowski spacetime. 
In Section~\ref{sec:proof2}, we consider a spacetime with non-vanishing Weyl tensor, and prove Theorem~\ref{th:Weyl}. The proof is based on constructing a concrete example of eight events that demonstrates violation of well-stitchedness for such a spacetime. 



\section{Concept of a well-stitched spacetime}
\label{sec:theorems}

Let us denote as $A\lightcone B$ the situation when event $A$ belongs to the light cone of event $B$ (or, equivalently, $B$ belongs to the light cone of $A$). In a flat 4-dimensional spacetime (i.~e. a Minkowski spacetime), $A\lightcone B$ means that
\begin{equation}
\label{eq:LC-Minkowski}
(x_A-x_B)^2 + (y_A-y_B)^2 + (z_A-z_B)^2 = c^2(t_A-t_B)^2 ,
\end{equation}
where $x$, $y$ and $z$ are Cartesian coordinates, $t$ is the time, and $c$ is the speed of light. In a curved (pseudo-Riemannian) spacetime, $A\lightcone B$ means that events $A$ and $B$ are connected by a light-like geodesic line.

This relation allows to express the conformal flatness of a 4-dimensional spacetime in a very simple way. Namely, we will show that conformal flatness is equivalent to ``well-stitchedness'' of a spacetime defined as follows:

\begin{definition}[well-stitched spacetime]
A four-dimensional spacetime is well-stitched iff for any eight events $A$, $B$, $C$, $D$, $E$, $F$, $G$ and $H$, that are all different from each other, from relations $A\lightcone E$, $A\lightcone F$, $A\lightcone G$, $A\lightcone H$, $B\lightcone E$, $B\lightcone F$, $B\lightcone G$, $B\lightcone H$, $C\lightcone E$, $C\lightcone F$, $C\lightcone G$, $C\lightcone H$, $D\lightcone E$, $D\lightcone F$ and $D\lightcone G$ follows relation $D\lightcone H$.
\label{def:1}
\end{definition}

Table \ref{table:1} clarifies this set or relations. The relation between $D$ and $H$, denoted by the question mark in the table, follows from other 15 relations denoted by symbols~$\lightcone$.
\begin{table}
\centering
\begin{tabular}{ c|cccc| } 
     & $A$ & $B$ & $C$ & $D$ \\ 
 \hline
 $E$ & $\lightcone $ & $\lightcone $ & $\lightcone $ & $\lightcone $ \\ 
 $F$ & $\lightcone $ & $\lightcone $ & $\lightcone $ & $\lightcone $ \\ 
 $G$ & $\lightcone $ & $\lightcone $ & $\lightcone $ & $\lightcone $ \\ 
 $H$ & $\lightcone $ & $\lightcone $ & $\lightcone $ & $?$ \\ 
 \hline
\end{tabular}
\caption{Illustration of the set of relations in Definition~\ref{def:1}.}
\label{table:1}
\end{table}

The thought experiment shown in Fig.~\ref{fig:gedanken}a,b is a specific case of the set of relations that appears in the definition of a well-stitched spacetime. In this experiment, as shown schematically in Fig.~\ref{fig:gedanken}c, events $E$, $F$, $G$, $H$ of switching lamps on precede events $A$, $B$, $C$, $D$ of their perceptions by the observers. Other timelines are also possible. For example, Fig.~\ref{fig:gedanken}d demonstrates another version of the thought experiment, in which two lamps are switched on at events $B$ and $D$; then the light signals are percepted at events $E$, $F$, $G$, $H$ and retranslated further from these four events; finally, retranslated signals reach events $A$ and $C$. 
The setting shown in Fig.~\ref{fig:gedanken}d will be used for the proof of Theorem~\ref{th:Weyl} in Section~\ref{sec:proof2}.

In order to establish the equivalence between well-stitchedness and conformal flatness, we will prove two theorems.

\begin{theorem}
The 4-dimensional Minkowski spacetime is well-stitched.
\label{th:Minkowski}
\end{theorem}

This theorem will be proven in Section~\ref{sec:proof1}. 

A conformal transformation~(\ref{eq:conformal}), which maps a conformally flat spacetime onto Minkowski one, conserves causal relations. Therefore it conserves well-stitchedness as well, because this property relies on causal relations only. Hence, Theorem~\ref{th:Minkowski} implies the following result:

\begin{corollary}
Any 4-dimensional conformally flat spacetime is well-stitched.
\label{cor:Minkowski}
\end{corollary}

Then, in order to close our reasoning, it is necessary to consider \emph{non}-conformally-flat spacetimes, and prove that they \emph{are not} well-stitched. In this way, the following theorem is helpful:
\begin{theorem}
If a 4-dimensional spacetime has a non-zero Weyl tensor at some point $O$, then well-stitchedness violates in some vicinity of $O$.
\label{th:Weyl}
\end{theorem}
The Weyl tensor here is a traceless part of the Riemann curvature tensor~\cite[\S92]{LL2}. We will prove this theorem in Section~\ref{sec:proof2}.

It is well known that a 4-dimensional spacetime is conformally flat if and only if its Weyl tensor is equal to zero everywhere~\cite[Sec.~6.13]{dInverno_book}. Let us consider a spacetime that is not conformally flat. According to the above-mentioned property, there is some point at which the Weyl tensor is different from zero. Therefore, Theorem~\ref{th:Weyl} entails the following 
\begin{corollary}
If a 4-dimensional spacetime is not conformally flat, then it is not well-stitched.
\label{cor:Weyl}
\end{corollary}

Corollaries \ref{cor:Minkowski} and \ref{cor:Weyl} together consist the main result of this paper:\newline
\textbf{A 4-dimensional spacetime is conformally flat if and only if it is well-stitched.}


\section{Possible implications}
\label{sec:discussion}

Let us discuss, whether it is possible in practice (rather than in theory) to detect the spacetime curvature produced by the Earth by finding out a violation of well-stitchedness. The curvature is described by the Riemann tensor $R_{iklm}$ that has 20 degrees of freedom. Ten of them are components of the Ricci tensor $R_{ik}$, and in an empty space they vanish according to Einstein equations. Other ten degrees of freedom are independent components of the Weyl tensor $C_{iklm}$. Moreover, in an empty space, where $R_{ik}=0$, the Riemann and Weyl tensors are equal to each other, see Eq.~(\ref{eq:2:Weyl-def}). Hence, in the empty space (e.~g. on an orbit around the Earth) a full characterization of the curvature is just the same as measuring the Weyl tensor.

\begin{figure}
\includegraphics[width=\linewidth]{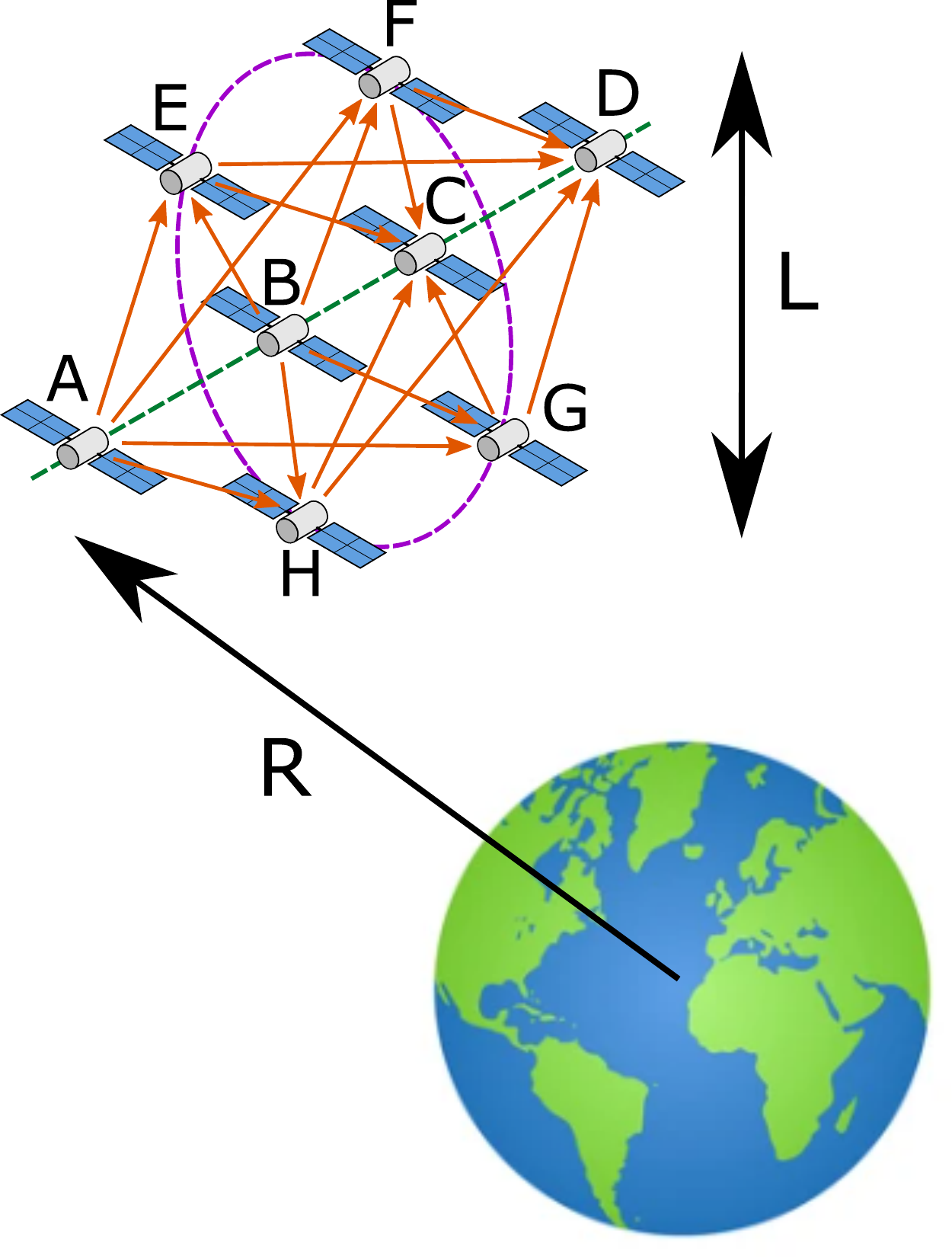}
\caption{Possible arrangement of a swarm of satellites near the Earth, enabling one to detect the curvature of the space due to the Earth's gravity. One group of satellites is located on a circle (violet dashed line), another one --- on the axis of symmetry of this circle (green dashed line). Yellow arrows show the directions of optical signals between the satellites. $L$ is the size of the swarm, and $R$ is the distance from the Earth's center.}
\label{fig:satellites}
\end{figure}

Consider a set of satellites orbiting around the Earth and forming a suitable configuration for detecting the spacetime curvature. An example of such a configuration is shown in Fig.~\ref{fig:satellites}. Satellite $A$ simultaneously sends optical signals towards satellites $E$, $F$, $G$ and $H$. Satellite $B$ does the same in another moment of time, so that each of satellites $E$, $F$, $G$ and $H$ receives signals from $A$ and from $B$ simultaneously. As soon as satellite $E$ receives the signals, it emits its own signals towards satellites $C$ and $D$, and so do satellites $F$, $G$ and $H$. Satellite $C$ is  positioned such that all four signals towards $C$ are received simultaneously. Similarly, satellite $D$ is set at the point where signals from $E$, $F$ and $G$ comes at the same time. Then, to reveal the spacetime curvature, one has to detect the difference in time $\delta t_{HD}$ between the moment when signals from $E$, $F$ and $G$ reach satellite $D$, and the moment when the signal from $H$ reaches there.

How large is the time difference $\delta t_{HD}$ that is needed to be detected? One can apply an estimate from Subsection~\ref{sec:delta-t} and find that
\begin{equation}
\label{eq:delta-t-via-C}
\delta t_{HD} \simeq \frac{L^3 C}{c} \, ,
\end{equation}
where $L$ is a characteristic distance between the satellites, $C$ is a suitable component (or a linear combination of components) of the Weyl tensor, and $c$ is the speed of light. If the circle $EFGH$ lies in the $yz$-plane, and satellites $A$, $B$, $C$ and $D$ are located at equal distances from the center of the circle, then the configuration of events $A,\ldots,H$ is the same as used in Section~\ref{sec:proof2}, up to relabeling. In these case, quantity $C$ in the latter equation is component $C_{tyxy}$, as follows from Subsection~\ref{sec:delta-t}. Some other configurations of events can be obtained from that considered in Section~\ref{sec:proof2} by Lorentz boosts and/or spatial rotations of the satellite group. They enable us to get access to other components of the Weyl tensor and, eventually, obtain the full information about the spacetime curvature by measuring time delays in 10 different configurations of satellites.

The Riemann tensor scales with the distance $R$ from the center of a massive body as $\simeq r_g/R^3$, where $r_g$ is the gravitational radius of the body (see ~\cite[\S100]{LL2} for the Riemann tensor in spherical coordinates). Substituting this expression as $C$ into Eq.~(\ref{eq:delta-t-via-C}), we obtain
\begin{equation}
\label{eq:delta-t-via-R}
\delta t_{HD} \simeq \frac{r_g}{c} \left(\frac{L}{R}\right)^3 \, .
\end{equation}
In the limiting case $L \simeq R$, which corresponds to a setting shown in Fig.~\ref{fig:gedanken}b, quantity $\delta t_{HD}$ turns into the Shapiro time delay that has the order of magnitude $\simeq r_g/c$. 

Near the Earth ($r_g \approx 1$ cm), Eq.~(\ref{eq:delta-t-via-R}) gives rise to an estimate $\delta t_{HD} \simeq 0.3\,{\rm ns} \times (L/R)^3$. Hence, detection of the curvature related to the Earth gravity demands picosecond resolution for optical signal detection, which is routinely accessed at present. The same level of resolution is necessary for adjusting positions of satellites by optical signaling, in order to fulfill fifteen relations $A\lightcone E$, $A\lightcone F$ and so on (or, equivalently, to set fifteen time delays $\delta t_{AE}$, $\delta t_{AF}$, etc. to zero). For a solar orbit, estimate~(\ref{eq:delta-t-via-R}) with the gravitational radius of the Sun $r_g \approx 3$ km provides the time delay $\delta t_{HD}$ of order of magnitude $10\,\mu{\rm s} \times (L/R)^3$.

It is important to note that, even in absence of exact positioning, one can get an information about spacetime curvature from sixteen time delays $\delta t_{AE}, \delta t_{AF}, \ldots, \delta t_{HD}$. It can be done by mere replacement of $\delta t_{HD}$ in Eq.~(\ref{eq:delta-t-via-C}) with a suitable linear combination of time delays $\delta t_{AE}, \delta t_{AF}, \ldots, \delta t_{HD}$, which is stable with respect to small variations of satellite positions. Existence of such a stable linear combination is established in Subsection~\ref{sec:stability} for a special configuration of events. 

We wish to emphasize two distinctive features of the proposed method of detecting the spacetime curvature. First, {\bf it provides a direct access to the local value of the curvature tensor}. This is in contrast to other tests of the general relativity, which deal with integrated characteristics such as Mercury orbit precession, light deflection by the Sun, and so on. Second, {\bf it does not demand precise instruments}, like atomic clocks, drag-free satellite control, etc.~\cite{Lasers_Clocks_book} To obtain the spacetime curvature by our method with some relative accuracy $\eta$, one have to measure time delays with the same order $\eta$ of relative accuracy.

Regardless the possibility of testing spacetime curvature in practice, the results of this study raise some fundamental issues. Theorems~\ref{th:Minkowski} and~\ref{th:Weyl} state that, for 4-dimensional pseudo-Riemannian spaces, conformal flatness is equivalent to well-stitchedness. Conformal flatness is a concept of differential geometry, whereas a new notion of well-stitchedness, introduced in Definition~\ref{def:1}, belongs to \emph{incidence geometry}. Hence, Theorems~\ref{th:Minkowski} and~\ref{th:Weyl} ``translate'' conformal flatness from the language of differential geometry to a more primitive language of incidence geometry. In the latter language, a set of events connected by null geodesic lines, as in Definition~\ref{def:1}, consists a configuration of type $(8_4 16_2)$, which means a collection of 8 points and 16 lines, with 4 lines at each point and 2 points at each line~\cite[Chapter~III]{Hilbert_book}. Especially significant are those configurations, in which the last incidence is self-fulfilling, as for example in Pappus $(9_3)$ and Desargues $(10_3)$ configurations. Configurations that have this property express some geometrical theorems~\cite{Hilbert_book}. The configuration $(8_4 16_2)$ related to in Definition~\ref{def:1} also has this property in the 4-dimensional Minkowski spacetime, that is expressed in Theorem~\ref{th:Minkowski}.

There is an approach to quantum gravity, in which the spacetime is discrete and is represented as a \emph{causal set} --- a collection of discrete points, partially ordered by causality relations~\cite{Dribus_book,Surya2019,Loll2020}. In this perspective, the concept of a well-stitched spacetime may provide a convenient tool for distinguishing between a (conformally) flat causal set and a curved one. The fraction of violations of well-stitchedness in a causal set may serve as a measure of the Weyl curvature (cf. a method of scalar curvature evaluation in Ref.~\cite{Benincasa2010}).

It is known since 1970s that the \emph{causal structure} of a pseudo-Riemannian spacetime is in one-to-one correspondence with its \emph{conformal structure}~\cite{Hawking1976,Malament1977}. That is, if a spacetime can be mapped to another one with conserving causal relations, then the map between the spacetimes is conformal (i.~e. obeys Eq.~(\ref{eq:conformal})), and vice versa~\cite[Sec.~2.8]{Dribus_book}. The present study reveals the primitive elements of which the causal structure of a conformally flat 4-dimensional spacetime consists---these elements are configurations $(8_4 16_2)$, such as ones shown in Fig.~\ref{fig:gedanken}. 

\begin{figure}
\includegraphics[width=7cm]{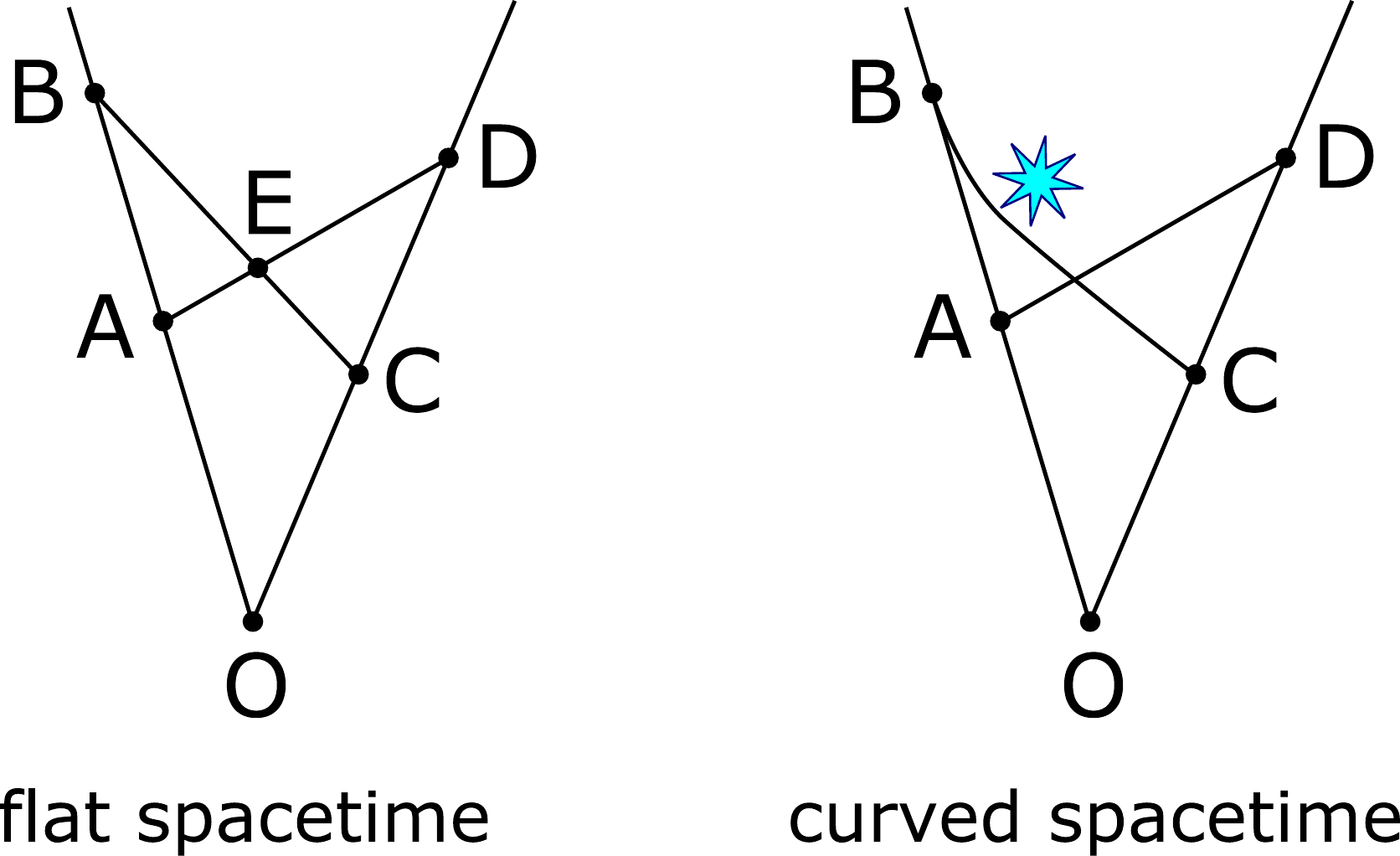}
\caption{A complete quadrilateral, or a configuration $(6_2 4_3)$, composed of four geodesic lines $OAB$, $OCD$, $AD$ and $BC$. 
In a flat spacetime, as well as in a spacetime of a constant curvature, lines $AD$ and $BC$ must intersect at some point $E$ (the left part). 
But in a generic curved spacetime, these lines may not intersect, for instance, if geodesic line BC passes near a massive body (the right part).}
\label{fig:quadrilateral}
\end{figure}

Another view to a spacetime without involving rulers and clocks is its \emph{geodesic (projective) structure}, i.~e. the set of geodesic lines, or in physical terms---the set of possible world lines of freely falling test particles~\cite{Topalov-Matveev2003,Matveev2012}. Remarkably, geodesic and conformal structures together determine the \emph{Weyl structure}~\cite{Ehlers2012,Matveev-Scholz2020}, that in the Riemannian geometry allows also to reconstruct the metric (e.~g. using the methods of Refs.~\cite{Desloge1989metrosphere,Desloge1989clock}). Spacetimes are called geodesically equivalent if they can be mapped into each other with preserving the geodesic lines. According to Beltrami's theorem~\cite[Sec.~40]{Eisenhart_book}, the flat Minkowski spacetime is geodesically equivalent to other spacetimes with a constant curvature, and only to them: namely, to de Sitter (positive curvature) and anti-de Sitter (negative curvature) spacetimes. In the context of the present study, it is natural to ask: whether it is possible to detect the curvature of a spacetime by pointing out to a finite set of geodesic lines? We will show that it is indeed possible, unless the spacetime has a constant curvature and therefore is geodesically equivalent to the flat one. Consider four geodesic lines $OAB$, $OCD$, $AD$ and $BC$ that form the so-called \emph{complete quadrilateral}, shown in Fig.~\ref{fig:quadrilateral}. Point $O$, two geodesic lines $a$ and $b$ passing through $O$, two points $A$, $B$ on line $a$, and two points $C$, $D$ on line $b$ are chosen arbitrarily, with the only restriction that point $A$ lies between $O$ and $B$, and point $C$ --- between $O$ and $D$. In the flat spacetime, where geodesics are just straight lines, five points $O$, $A$, $B$, $C$ and $D$ lie in one plane, and therefore lines $AD$ and $BC$ must intersect at some point $E$. The same argument is true in a spacetime of a constant curvature. But in a general curved spacetime lines $AD$ and $BC$ may not intersect. This occurs, for example, if geodesic line $BC$ passes near a massive body that deflects this line out of plane $OABCD$. Therefore spacetime curvature can be detected by the observation that lines $AD$ and $BC$ do not intersect each other.

We will show in Appendix~\ref{app:four-geodesics} that the distance between lines $AD$ and $BC$, projected to a vector perpendicular to both lines, is proportional to the cube of the quadrilateral's size multiplied by a linear combination of the Riemann tensor components. Considering different such quadrilaterals around point $O$, one can measure 19 of 20 degrees of freedom of the Riemann tensor at point $O$: namely, 10 independent components of the Weyl tensor $C_{iklm}$ and 9 components of the traceless part $(R_{ik} - \frac14 R{^l}_l \, g_{ik})$ of the Ricci tensor $R_{ik}$ (see Appendix~\ref{app:four-geodesics}). If in any quadrilateral $OABCD$ its lines $AD$ and $BC$ intersect, then the above-mentioned 19 degrees of freedom vanish at each point, i.~e. the spacetime is both conformally-flat ($C_{iklm}=0$) and Einstein ($R_{ik} = \frac14 R{^l}_l \, g_{ik}$). It follows from these properties~\cite[Sec.~5.2]{Hawking_book} that the spacetime has a constant curvature, i.~e. is either flat or de Sitter or anti-de Sitter. On the other hand, if in some such quadrilateral the lines $AD$ and $BC$ do not intersect, then the spacetime is not of a constant curvature. Hence, the finite set of geodesic lines shown in Fig.~\ref{fig:quadrilateral} allows us to discriminate between spacetimes of a constant curvature and all other spacetimes.

\begin{figure}
\includegraphics[width=5cm]{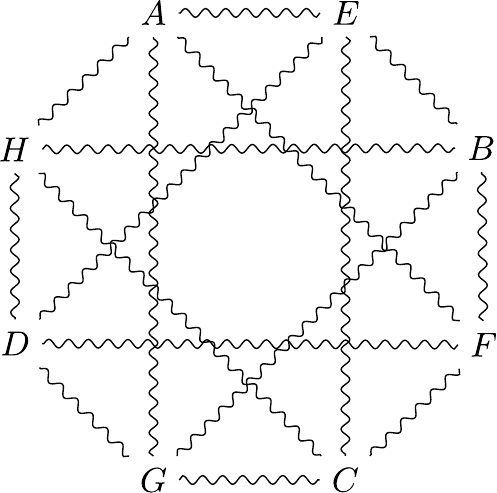}
\caption{An eight-vertex Feynman diagram that may describe a correction to the vacuum energy sensitive to the Weyl tensor.}
\label{fig:diagram}
\end{figure}

In the quantum realm, geodesic lines loose their physical meaning of trajectories of freely falling particles. But the fundamental role of causal relations still persists in quantum theory. One can thus expect that the property of well-stitchedness may be important in the context of quantum field theory in a curved spacetime. As an illustrative example, let us consider a Feynman diagram shown in Fig.~\ref{fig:diagram}, where sixteen lines connect eight vertices $A,\ldots,H$ in the same manner as causal relations connect eight events in Definition~\ref{def:1}. Such a diagram may appear in a quantum field with a four-particle interaction, for example in the $\varphi^4$-model, and represent a correction to the vacuum energy. 
Remarkably, gravity-induced four-\emph{fermion} interaction naturally appears in the framework of Einstein-Cartan theory with torsion~\cite{Khriplovich2013,Magueijo2013,Boos2017}.

The propagators in the position space, related to lines $AE$, $AF$, ..., and $DH$, diverge when the corresponding relations $A\lightcone E$, $A\lightcone F$, ..., and $D\lightcone H$ are satisfied. Therefore, in a well-stitched spacetime, divergence of fifteen propagators leads to divergence of the rest (sixteenth) propagator. In a not-well-stitched spacetime it is not the case. One may thus expect that the contribution $\Delta\mathcal L$ of the diagram shown in Fig.~\ref{fig:diagram} to the vacuum Lagrangian density would depend on whether the spacetime is well-stitched. On the other hand, presence or absence of well-stitchedness is determined by Weyl tensor $C_{iklm}$, due to Theorems~\ref{th:Minkowski} and~\ref{th:Weyl}. Hence, it is natural to suppose that contribution $\Delta\mathcal L$ is a function of $C_{iklm}$. Considering the expansion of $\Delta\mathcal L$ in a power series over components of the Weyl tensor, one can easily see that linear terms vanish because there is no scalar linear combination of components $C_{iklm}$. One can however construct of the Weyl tensor a scalar quadratic form, namely $C^{iklm} C_{iklm}$. For this reason, one can expect that the first non-vanishing curvature-dependent term in $\Delta\mathcal L$ is quadratic in $C_{iklm}$:
\begin{equation}
\label{eq:Weyl-action-1}
\Delta\mathcal L (C_{iklm}) - \Delta\mathcal L (0) \propto C^{iklm} C_{iklm} \sqrt{-g} ,
\end{equation}
where factor $\sqrt{-g}$ accounts for the invariant volume element, $g$ being the determinant of the metric tensor. That is, the curvature-related action $S_C$ that arises from the diagram shown in Fig.~\ref{fig:diagram} may be of the following form:
\begin{equation}
\label{eq:Weyl-action-2}
S_C = \alpha_g \int C^{iklm} C_{iklm} \sqrt{-g} \, {\rm d}t \, {\rm d}x \, {\rm d}y \, {\rm d}z ,
\end{equation}
where $\alpha_g$ is some constant. In a system of units where $\hbar=c=1$, $\alpha_g$ is a dimensionless quantity. The strength of the four-boson interaction $\lambda$ is also dimensionless, and it enters into $\alpha_g$ as $\lambda^8$ according to the number of vertices in the diagram. Hence $\alpha_g \simeq \lambda^8$ up to a numerical factor. In physical units, the latter relation has a form
\begin{equation}
\label{eq:Weyl-action-3}
\alpha_g \simeq \lambda^8 \hbar c .
\end{equation}
Remarkably, $\alpha_g$ does not depend on the particle's mass. Of course, a thorough calculation of constant $\alpha_g$ must take renormalization of $\lambda$ into account.

Action $S_C$ in Eq.~(\ref{eq:Weyl-action-2}) is known as Weyl action. It consists the basis of so-called conformal gravity theory. Though Weyl action substantially differs from Einstein-Hilbert action $S_{EH}$ of general relativity, $S_{EH} \propto \int R \sqrt{-g} \, {\rm d}t \, {\rm d}x \, {\rm d}y \, {\rm d}z$ (where $R$ is the scalar curvature), there are intriguing relations between conformal gravity and standard Einstein's general relativity. In particular, conformal gravity appeared in the context of supergravity~\cite{Fradkin1985}, twistor-string theory~\cite{Berkovits2004}, ultraviolet regularization of gravity~\cite{Avramidy1985}, astrophysical observations~\cite{Mannheim2012,Yang2013} as an origin of Planck and electroweak scales~\cite{Oda2018}, as a more fundamental theory beyond Einstein's gravity~\cite{Zee1983,Maldacena2011,Anastasiou2016,Anastasiou2021}, and even as a candidate alternative to Einstein gravity~\cite{Mannheim2012}. Conformal gravity also considered to induce a Starobinsky-type cosmological inflation~\cite{Jizba2015}. 

The idea that the action of the gravitational field stems from dynamics of the matter in a curved spacetime was formulated by Sakharov as early as in 1967 (see reprints~\cite{Sakharov1991,Sakharov2000} of his 1967 work). This idea is often referred as ``induced gravity''~\cite{Adler1982,Visser2002,Altshuler2021}. In a closely related concept of ``emergent gravity'', gravitational field arises along with spacetime itself from some more fundamental constructs~\cite{Konopka2006,Dribus_book,Verlinde2011,Sindoni2012,Berman2022}. Calculations by many research groups showed that the induced gravity appears even in the approximation of non-interacting quantum fields, from a diagram containing only one loop~\cite{Sakharov1975,Christensen1980,Adler1982,Birrell_Davies_book,Visser2002,Kehagias2021,Altshuler2021}. However, a much more complicated diagram shown in Fig.~\ref{fig:diagram} might be interesting in the aspect that it naturally leads to the induced \emph{conformal} gravity.


\section*{Conclusions}
\label{sec:conclusions}

In this work, we have found a \emph{discrete} equivalent of a conformally flat 4-dimensional pseudo-Riemannian spacetime. This equivalent is a new notion of a ``well-stitched'' spacetime, inspired by the thought experiment depicted in Fig.~\ref{fig:gedanken}, and further developed in Section~\ref{sec:theorems}. The well-stitched spacetime is defined solely in terms of light signals between a finite number of events (see Definition~\ref{def:1}), without using rulers or clocks of any sort.

We have proved two theorems about well-stitched spacetimes. Theorem~\ref{th:Minkowski} states that any conformally flat 4-dimensional pseudo-Riemannian spacetime is well-stitched. Theorem~\ref{th:Weyl}, conversely, states that any \emph{non}-conformally-flat 4-dimensional pseudo-Riemannian spacetime is \emph{non}-well-stitched. Hence, the differential-geometric concept of conformal flatness has been ``translated'' to the language of discrete geometry. 

This new look onto the spacetime geometry opens different perspectives (see Section~\ref{sec:discussion}), ranging from the curvature detection by time-delay measurements with satellites to the search for an origin of gravity.

\begin{acknowledgments}
A.N. thanks the Faculty of Physics of the Philipps Universit\"at Marburg for the kind hospitality during his research stay.
\end{acknowledgments}

\newpage

\section*{Methods}

\section{Causal structure of the flat spacetime: proof of Theorem~\ref{th:Minkowski}}
\label{sec:proof1}

In this Section, we consider points (events) in the 4-dimensional Minkowski space. For simplicity, we set speed of light $c$ equal to 1. The interval $ds^2_{PQ}$ between two points $P = (t_P,x_P,y_P,z_P)$ and $Q = (t_Q,x_Q,y_Q,z_Q)$ is defined as
\begin{equation}
\label{eq:1:interval}
ds_{PQ}^2 = -(t_P-t_Q)^2 + (x_P-x_Q)^2 + (y_P-y_Q)^2 + (z_P-z_Q)^2,
\end{equation}
and relation $P\lightcone Q$ just means that $ds_{PQ}^2 = 0$. The scalar product of two contravariant vectors $\vec{a} = (a_t,a_x,a_y,a_z)$ and $\vec{b} = (b_t,b_x,b_y,b_z)$ is 
\begin{equation}
\label{eq:1:scalar_product}
\vec{a} \cdot \vec{b} = -a_tb_t + a_xb_x + a_yb_y + a_zb_z \, .
\end{equation}

Our proof of Theorem~\ref{th:Minkowski} is based on three lemmas formulated below.
\begin{lemma}
\label{lemma:scalar-product}
For any four points $P$, $Q$, $R$, $S$, if relations $P\lightcone R$, $P\lightcone S$, $Q\lightcone R$ and $Q\lightcone S$ are fulfilled, then vector $\overrightarrow{PQ}$ is perpendicular to vector $\overrightarrow{RS}$ (that is, $\overrightarrow{PQ} \cdot \overrightarrow{RS} = 0$).
\end{lemma}
\begin{lemma}
\label{lemma:one-line}
For any three different points $P$, $Q$, $R$, if relations $P\lightcone Q$, $P\lightcone R$ and $Q\lightcone R$ are fulfilled, then these points lie on one line.
\end{lemma}
\begin{lemma}
\label{lemma:light-like}
For any four points $P$, $Q$, $R$, $S$, if relations $P\lightcone S$, $Q\lightcone S$ and $R\lightcone S$ are fulfilled, and points $P$, $Q$, $R$ lie on one line and are different from each other, then this line is light-like (that is, $P\lightcone Q$, $P\lightcone R$ and $Q\lightcone R$).
\end{lemma}
These lemmas are true in the Minkowski space. Proofs of them are given in Appendices~\ref{app:proof-scalar-product}, \ref{app:proof-one-line} and~\ref{app:proof-light-like}.

Now we return to Theorem~\ref{th:Minkowski}. Let eight points $A$, $B$, $C$, $D$, $E$, $F$, $G$, $H$ be all different from each other, and 15 relations $A\lightcone E, \ldots, D\lightcone G$ listed in Definition~\ref{def:1} are fulfilled. 
First, we consider the generic case, when points $A$, $B$ and $C$ do not lie on one line, and also points $E$, $F$ and $G$ do not lie on one line. The opposite case will be discussed later.

Due to Lemma~\ref{lemma:scalar-product}, $\overrightarrow{AB} \cdot \overrightarrow{EF} = 0$, $\overrightarrow{AB} \cdot \overrightarrow{EG} = 0$, $\overrightarrow{AC} \cdot \overrightarrow{EF} = 0$, and $\overrightarrow{AC} \cdot \overrightarrow{EG} = 0$. Therefore any vector parallel to plane $ABC$ is perpendicular to any vector parallel to plane $EFG$.

Since $A\lightcone E$, $A\lightcone F$, $D\lightcone E$ and $D\lightcone F$, it follows from Lemma~\ref{lemma:scalar-product} that vector $\overrightarrow{AD}$ is perpendicular to vector $\overrightarrow{EF}$. Similarly, as a consequence of relations $A\lightcone E$, $A\lightcone G$, $D\lightcone E$ and $D\lightcone G$, vector $\overrightarrow{AD}$ is perpendicular to $\overrightarrow{EG}$. Therefore, vector $\overrightarrow{AD}$ is perpendicular to plane $EFG$ 
and hence is parallel to plane $ABC$. 
This means that points $A$, $B$, $C$ and $D$ lie on one plane. Analogous reasoning shows that points $E$, $F$, $G$ and $H$ lie on one plane.

Hence, all the eight points lie in two planes $ABCD$ and $EFGH$, and each vector of the first plane is perpendicular to each vector in the second one.

There are three options: (a) plane $ABCD$ contains a time-like vector; (b) plane $EFGH$ contains a time-like vector; (c) none of these planes contains a time-like vector. We will consider them separately.

In option (a), we can choose axis $t$ along a time-like vector lying in plane $ABCD$. Then, any vector in plane $EFGH$ is perpendicular to axis $t$. Let us choose axes $y$ and $z$ along some pair of mutually perpendicular vectors of plane $EFGH$. Axis $x$ is then perpendicular to plane $EFGH$ 
and therefore must be parallel to plane $ABCD$. 
Planes $ABCD$ and $EFGH$ intersect at some point $O$. Let us choose this point as the origin of the coordinate axes.

\begin{figure}
\includegraphics[width=7cm]{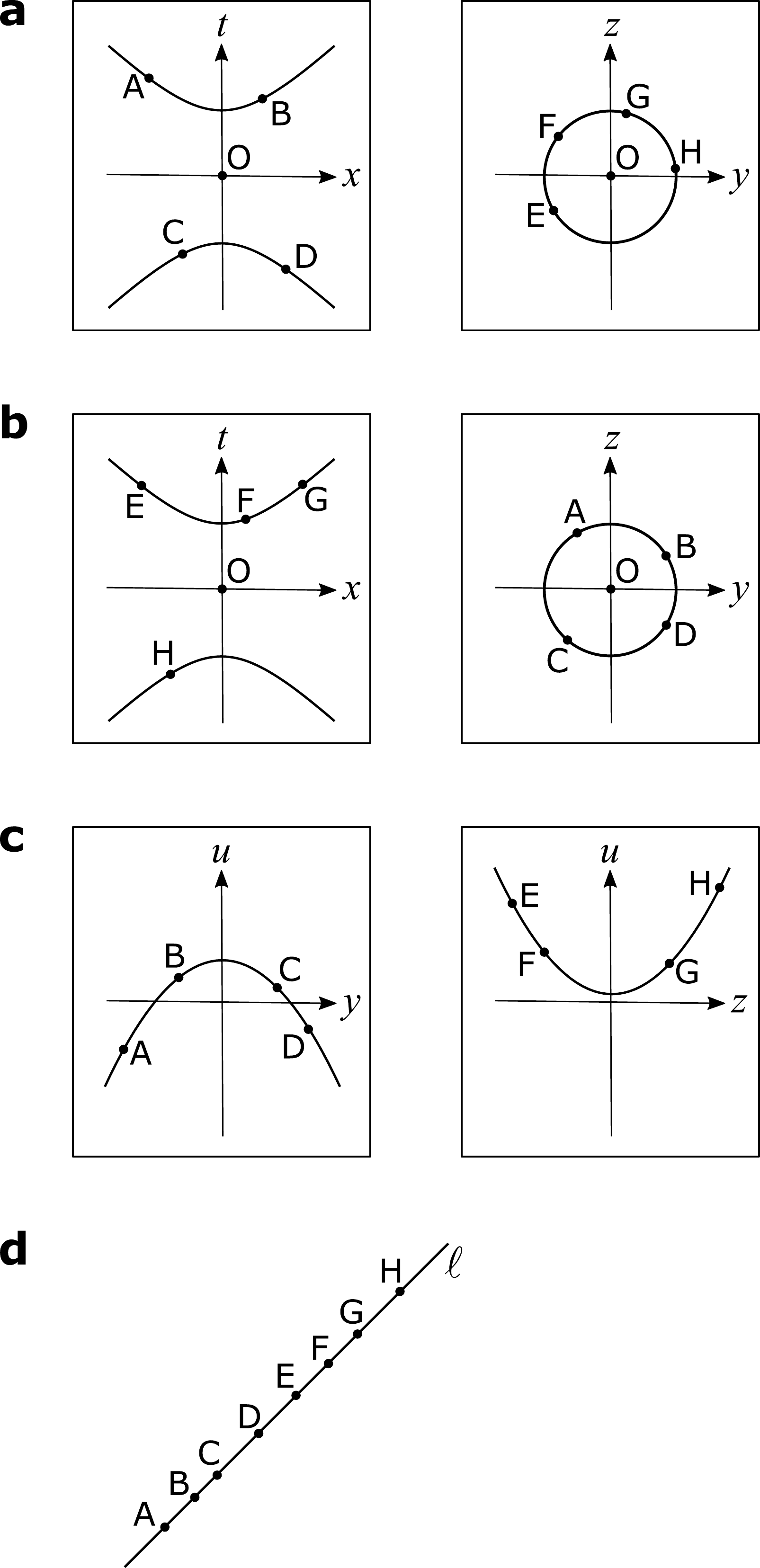}
\caption{Four possible layouts of points (events) $A$, ..., $H$ in the flat spacetime, shown in specially chosen coordinate frames: 
(a) points $A$, $B$, $C$, $D$ lie on a hyperbola in plane $xt$, and points $E$, $F$, $G$, $H$ --- on a circle in plane $yz$; 
(b) points $E$, $F$, $G$, $H$ lie on a hyperbola in plane $xt$, and points $A$, $B$, $C$, $D$ --- on a circle in plane $yz$; 
(c) points $A$, $B$, $C$, $D$ lie on a parabola in plane $t-x=0$, $z=0$, and points $E$, $F$, $G$, $H$ --- on a parabola in plane $t-x=s$, $y=0$; 
(d) all eight points lie on the same light-like line $\ell$. 
Point $O$ is the intersection point of planes $ABCD$ and $EFGH$. Axis $u$ is a bisector of the angle between axes $x$ and $t$.}
\label{fig:points}
\end{figure}

With this choice of coordinates, points $A$, $B$, $C$ and $D$ lie in the plane spanned onto axes $t$ and $x$. Similarly, points $E$, $F$, $G$ and $H$ lie in the plane spanned onto axes $y$ and $z$. That is,
\begin{gather}
y_A = z_A = 0, \\
y_B = z_B = 0, \\
y_C = z_C = 0, \\
y_D = z_D = 0, \\
t_E = x_E = 0, \\
t_F = x_F = 0, \\
t_G = x_G = 0, \\
t_H = x_H = 0.
\end{gather}
Condition $A\lightcone E$ therefore acquires the form
\begin{equation}
\label{eq:1:AE}
-t_A^2 + x_A^2 + y_E^2 + z_E^2 = 0.
\end{equation}
Let us denote the distance from point $E$ to the origin of coordinates as $r$:
\begin{equation}
\label{eq:1:E-r}
y_E^2 + z_E^2 = r^2.
\end{equation}
Then, it follows from Eqs.~(\ref{eq:1:AE}) and~(\ref{eq:1:E-r}) that
\begin{equation}
\label{eq:1:A-r}
t_A^2 - x_A^2 = r^2.
\end{equation}
Similarly, conditions $B\lightcone E$, $C\lightcone E$, $D\lightcone E$ along with Eq.~(\ref{eq:1:E-r}) lead to
\begin{gather}
\label{eq:1:B-r}
t_B^2 - x_B^2 = r^2, \\
\label{eq:1:C-r}
t_C^2 - x_C^2 = r^2, \\
\label{eq:1:D-r}
t_D^2 - x_D^2 = r^2, 
\end{gather}
and conditions $A\lightcone F$, $A\lightcone G$, $A\lightcone H$ and Eq.~(\ref{eq:1:A-r}) lead to
\begin{gather}
\label{eq:1:F-r}
y_F^2 + z_F^2 = r^2, \\
\label{eq:1:G-r}
y_G^2 + z_G^2 = r^2, \\
\label{eq:1:H-r}
y_H^2 + z_H^2 = r^2.
\end{gather}
Summarizing, we see that points $A$, $B$, $C$ and $D$ lie on the hyperbola
\begin{equation}
\label{eq:1:ABCD-r}
t^2 - x^2 = r^2, \quad y=0, \quad z=0,
\end{equation}
and points $E$, $F$, $G$ and $H$ lie on the circle
\begin{equation}
\label{eq:1:EFGH-r}
y^2 + z^2 = r^2, \quad t=0, \quad x=0.
\end{equation}
Such an arrangement of points is shown in Fig.~\ref{fig:points}a.

Finally, Eqs.~(\ref{eq:1:D-r}) and~(\ref{eq:1:H-r}) can be combined into
\begin{equation}
\label{eq:1:DH}
-t_D^2 + x_D^2 + y_H^2 + z_H^2 = 0,
\end{equation}
that means $D\lightcone H$. Hence, within option (a), Theorem~\ref{th:Minkowski} is proven.

Option (b) can be reduced to option (a) by mere relabeling $A \leftrightarrow E$, $B \leftrightarrow F$, $C \leftrightarrow G$, $D \leftrightarrow H$, see Fig.~\ref{fig:points}b. Thus the above proof is valid also in option (b).

In option (c), as we show in Appendix~\ref{app:option-c}, one can choose the coordinate system in such a way that plane $ABCD$ is defined as
\begin{equation}
\label{eq:1:ABCD-plane-option-c}
t - x = 0, \quad z = 0,
\end{equation}
and plane $EFGH$ is defined as
\begin{equation}
\label{eq:1:EFGH-plane-option-c}
t - x = s, \quad y = 0,
\end{equation}
with some parameter $s$. 
In these coordinates, relation $P\lightcone Q$, where $P\in\{A,B,C,D\}$ and $Q\in\{E,F,G,H\}$, has the form
\begin{equation}
\label{eq:1:PQ-option-c-0}
-(t_P-t_Q)^2 + (x_P-x_Q)^2 + y_P^2 + z_Q^2 = 0.
\end{equation}
Expressing here $t_P$ through $x_P$ by Eq.~(\ref{eq:1:ABCD-plane-option-c}), and $t_Q$ through $x_Q$ by Eq.~(\ref{eq:1:EFGH-plane-option-c}), one obtains
\begin{equation}
\label{eq:1:PQ-option-c}
2sx_P + y_P^2 = 2sx_Q - z_Q^2 + s^2.
\end{equation}
Let us denote as $w$ the following quantity:
\begin{equation}
\label{eq:1:A-w}
2sx_A + y_A^2 = w.
\end{equation}
Using Eq.~(\ref{eq:1:PQ-option-c}), one can easily deduce from relations $A\lightcone E$ and $A\lightcone H$ and Eq.~(\ref{eq:1:A-w}) that
\begin{gather}
\label{eq:1:E-w}
2sx_E - z_E^2 + s^2 = w, \\
\label{eq:1:H-w}
2sx_H - z_H^2 + s^2 = w.
\end{gather}
Then, it follows from relation $D\lightcone E$ and Eq.~(\ref{eq:1:E-w}) that
\begin{equation}
\label{eq:1:D-w}
2sx_D + y_D^2 = w.
\end{equation}
Points $A$, $B$, $C$, $D$ lie on a parabola $2sx + y^2 = w$, and points $E$, $F$, $G$, $H$ --- on a parabola $2sx - z^2 + s^2 = w$, see Fig.~\ref{fig:points}c.

Left-hand sides of Eqs.~(\ref{eq:1:D-w}) and~(\ref{eq:1:H-w}) are equal to each other:
\begin{equation}
\label{eq:1:DH-option-c}
2sx_D + y_D^2 = 2sx_H - z_H^2 + s^2.
\end{equation}
According to Eq.~(\ref{eq:1:PQ-option-c}), this means that relation $D\lightcone H$ is fulfilled in option (c). Q.E.D.

We have completed the proof in the generic case, when points $A$, $B$ and $C$ do not lie on one line, and also points $E$, $F$ and $G$ also do not lie on one line. 
What remains is to prove Theorem~\ref{th:Minkowski} in the opposite case. Let, for definiteness, points $A$, $B$ and $C$ lie on the same line $\ell$. 
Applying Lemma~\ref{lemma:light-like} with $P=A$, $Q=B$, $R=C$ and $S=E$, one can ensure that line $\ell$ is light-like, and that relations $A\lightcone B$, $A\lightcone C$ and $B\lightcone C$ are satisfied. Applying Lemma~\ref{lemma:one-line} five times: with $P=A$, $Q=B$, $R=E$, with $P=A$, $Q=B$, $R=F$, with $P=A$, $Q=B$, $R=G$, with $P=A$, $Q=B$, $R=H$, and then with $P=D$, $Q=E$, $R=F$, one can see that all eight points $A,B,C,D,E,F,G$ and $H$ lie on the same light-like line $\ell$ (see Fig.~\ref{fig:points}d). Since points $D$ and $H$ are connected by a light-like line, then relation $D\lightcone H$ is fulfilled. This completes the proof of Theorem~\ref{th:Minkowski}.


\section{Causal structure of a spacetime with nonzero Weyl tensor: proof of Theorem~\ref{th:Weyl}}
\label{sec:proof2}

According to the premise of Theorem~\ref{th:Weyl}, we suppose that Weyl tensor differs from zero at some point $O$. 
The Weyl tensor $C_{iklm}$ is a traceless part of Riemann curvature tensor~$R_{iklm}$. By definition, Weyl tensor is equal to~\cite[\S92]{LL2}
\begin{multline}
\label{eq:2:Weyl-def}
C_{iklm} = R_{iklm} - \frac12 R_{il}\, g_{km} + \frac12 R_{im}\, g_{kl} \\
+ \frac12 R_{kl}\, g_{im} - \frac12 R_{km}\, g_{il} 
+ \frac R6 \left( g_{il}\, g_{km} - g_{im}\, g_{kl} \right) ,
\end{multline}
where $R_{ik} = {R^l}_{ilk}$ is the Ricci tensor, and $R = {R^i}_i$ is the scalar curvature.

For convenience, we set the speed of light $c$ to 1 in this section. The signature adopted for the metric tensor is $(-+++)$.

\subsection{Choice of a reference frame}
\label{sec:frame}

In this section, we will use so-called \emph{Riemann normal coordinates}~\cite[\S11.6]{Wheeler_book} with the origin at point $O$. In these coordinates, metric tensor $g_{ik}$ at the origin is the same as that of the special relativity, and the first derivatives of $g_{ik}$ vanish at the same point:
\begin{gather}
\label{eq:2:g-is-eta}
g_{ik} = \eta_{ik} \text{ at } O, \\
\label{eq:2:g-deriv-zero}
\frac{\partial g_{ik}}{\partial r^l} = 0 \text{ at } O, 
\end{gather}
where
\begin{equation}
\label{eq:2:eta}
\eta_{ik} = 
\begin{pmatrix}
-1 & 0 & 0 & 0 \\
 0 & 1 & 0 & 0 \\
 0 & 0 & 1 & 0 \\
 0 & 0 & 0 & 1 
\end{pmatrix}
.
\end{equation}
The second derivatives of $g_{ik}$ are determined by the curvature tensor $R_{iklm}$, such that~\cite{Brewin2009}
\begin{equation}
\label{eq:2:g-series}
g_{ik}(\r) = \eta_{ik} - \frac13 R_{ilkm}(O) \, r^l r^m + o\left(|\r|^2\right),
\end{equation}
where $o(...)$ is the Landau little-o symbol, and $|\r|^2 = t^2+x^2+y^2+z^2$.

Additionally, we demand that
\begin{equation}
\label{eq:2:C-component-nonzero}
C_{tyxy}(O) \neq 0.
\end{equation}
If the latter inequality does not hold, then we perform an appropriate Lorentz transformation of the coordinates, after which inequality~(\ref{eq:2:C-component-nonzero}) becomes satisfied. The existence of such a transformation is proven in Appendix~\ref{app:C-inequality}.

\subsection{Intervals in a neighborhood of point $O$}
\label{sec:intervals}

In a flat spacetime, relation $P\lightcone Q$ means that the interval between points $P$ and $Q$ is equal to zero. 
If a spacetime is curved, the algebraic definition of the interval, as in Eq.~(\ref{eq:1:interval}), becomes meaningless. 
However, the interval can be determined by integration of the line element $(g_{ik} dr^i dr^k)^{1/2}$ along the geodesic line that connects two given points. 
Let us define interval $\delta s^2_{PQ}$ between points $P$ and $Q$ as
\begin{equation}
\label{eq:2:delta-s-def}
\delta s^2_{PQ} = \pm \left[ \int_P^Q \sqrt{\pm g_{ik} \frac{dr^i}{d\lambda} \, \frac{dr^k}{d\lambda}  } \;  d\lambda  \right]^2 ,
\end{equation}
where integration is performed over the geodesic line segment between $P$ and $Q$, $\lambda$ is a parameter along this geodesic line, $r^i(\lambda)$ are coordinates of a point on the geodesic line, and two signs $\pm$ are equal to each other and chosen such that the expression under the square root is non-negative. 

The function $\delta s^2_{PQ}$ defined by Eq.~(\ref{eq:2:delta-s-def}) is nothing else but Synge's world function $\Omega(PQ)$~\cite{Synge_book} multiplied by 2. If the geodesic line that connects $P$ and $Q$ is timelike, than $\delta s^2_{PQ} = -(\tau_{PQ})^2$, where $\tau_{PQ}$ is the proper time passed between events $P$ and $Q$ along the geodesic. If this geodesic line is spacelike, than $\delta s^2_{PQ}$ is the squared length of the geodesic line segment between $P$ and $Q$. And if this geodesic line is lightlike, than $\delta s^2_{PQ}=0$.

Hence, $\delta s^2_{PQ} = 0$ if and only if relation $P\lightcone Q$ is fulfilled.

In the Minkowski space, where $g_{ik} \equiv \eta_{ik}$, quantity $\delta s^2_{PQ}$ is the usual special-relativity interval:
\begin{equation}
\label{eq:2:delta-s-Minkowski}
\delta s^2_{PQ} = \eta_{ik} (r_Q^i - r_P^i) (r_Q^k - r_P^k) . \quad \text{(flat spacetime)}
\end{equation}
In a curved spacetime that is described by Eq.~(\ref{eq:2:g-series}), there must be corrections to Eq.~(\ref{eq:2:delta-s-Minkowski}). 
Let us define $\epsilon$-neighborhood of point $O$ as a set of such points $(t,x,y,z)$ that
\begin{equation}
\label{eq:2:neighborhood}
t^2 + x^2 + y^2 + z^2 < (2\epsilon)^2 .
\end{equation}
If both points $P$ and $Q$ belong to the $\epsilon$-neighborhood of point $O$, then~\cite{Brewin2009}
\begin{multline}
\label{eq:2:delta-s-via-A}
\delta s^2_{PQ} = \eta_{ik} (r_Q^i - r_P^i) (r_Q^k - r_P^k) \\
- \frac13 R_{iklm}(O)  \, r_P^i \, (r_Q^k - r_P^k)\, r_P^l (r_Q^m - r_P^m) + o(\epsilon^4).
\end{multline}
Later, we will use expansion~(\ref{eq:2:delta-s-via-A}) in order to construct a set of points that demonstrates violation of well-stitchedness.

\subsection{Combination of intervals and Weyl tensor}
\label{sec:combination}

In this subsection, we will establish a relation between intervals $\delta s^2_{PQ}$, on the one side, and the Weyl tensor, on the other side. Let us first express component $C_{tyxy}$ of the Weyl tensor at point $O$ through the Riemann curvature tensor $R_{iklm}$, using Eqs.~(\ref{eq:2:Weyl-def}), (\ref{eq:2:g-is-eta}), and the skew symmetry properties $R_{iklm} = -R_{kilm} = -R_{ikml}$:
\begin{equation}
\label{eq:2:C-tyxy-via-R}
C_{tyxy}(O) = \frac{ R_{tyxy}(O) - R_{tzxz}(O) }{2} \, .
\end{equation}

\begin{figure}
\includegraphics[width=\linewidth]{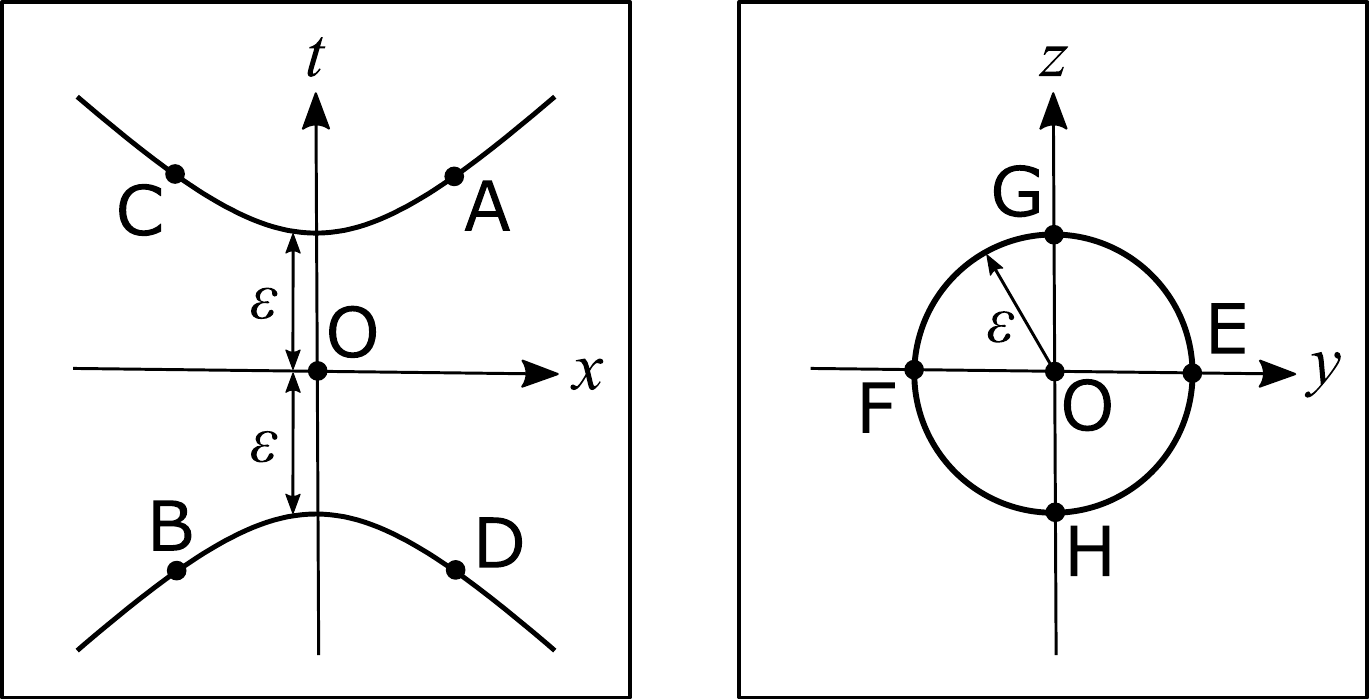}
\caption{Set of points (events) $A$, ..., $H$ defined by Eqs.~(\ref{eq:2:AH-def}).}
\label{fig:points2}
\end{figure}

Then we choose a spatial scale $\epsilon$, small enough such that expansion~(\ref{eq:2:delta-s-via-A}) makes sense, and consider a typical example of eight points (events) $A$ -- $H$ that obey sixteen relations $A\lightcone E, \ldots, D\lightcone H$ in the Minkowski space:
\begin{subequations} \label{eq:2:AH-def}
\begin{eqnarray}
A&=&(\;\;\,\sqrt2\epsilon,  \;\;\,\epsilon,  \;\;\,0,  \;\;\,0), \label{eq:2:A-def} \\
B&=&(-\sqrt2\epsilon,  -\epsilon,  \;\;\,0,  \;\;\,0), \label{eq:2:B-def} \\
C&=&(\;\;\,\sqrt2\epsilon,  -\epsilon,  \;\;\,0,  \;\;\,0), \label{eq:2:C-def} \\
D&=&(-\sqrt2\epsilon,  \;\;\,\epsilon,  \;\;\,0,  \;\;\,0), \label{eq:2:D-def} \\
E&=&(\;\;\;\;\;0,  \;\;\;\;\;0,  \;\;\,\epsilon,  \;\;\,0), \label{eq:2:E-def} \\
F&=&(\;\;\;\;\;0,  \;\;\;\;\;0,  -\epsilon,  \;\;\,0), \label{eq:2:F-def} \\
G&=&(\;\;\;\;\;0,  \;\;\;\;\;0,  \;\;\,0,  \;\;\,\epsilon), \label{eq:2:G-def} \\
H&=&(\;\;\;\;\;0,  \;\;\;\;\;0,  \;\;\,0,  -\epsilon), \label{eq:2:H-def} 
\end{eqnarray}
\end{subequations}
where coordinates of the points are listed in order $(t,x,y,z)$. 
This set of points is depicted in Fig.~\ref{fig:points2}, and is a particular case of the arrangement shown in Fig.~\ref{fig:points}a. 

Points $A$, $B$, $C$, and $D$ lie on hyperbola
\begin{equation}
\label{eq:2:ABCD-hyperbola}
t^2 - x^2 = \epsilon^2, \quad y=z=0,
\end{equation}
whereas points $E$, $F$, $G$, and $H$ lie on circle
\begin{equation}
\label{eq:2:EFGH-circle}
y^2 + z^2 = \epsilon^2, \quad t=x=0,
\end{equation}
in accordance with Eqs.~(\ref{eq:1:ABCD-r}) and~(\ref{eq:1:EFGH-r}) of Section~\ref{sec:proof1}. 
Let $P$ be one of points $\{A,B,C,D\}$, and $Q$ be one of points $\{E,F,G,H\}$. Then, due to Eqs.~(\ref{eq:2:ABCD-hyperbola}) and~(\ref{eq:2:EFGH-circle}),
\begin{multline}
\label{eq:2:PQ-interval-0}
\eta_{ik} (r_Q^i - r_P^i) (r_Q^k - r_P^k) \\
\equiv -(t_Q-t_P)^2 + (x_Q-x_P)^2 + (y_Q-y_P)^2 + (z_Q-z_P)^2 \\
= -t_P^2 + x_P^2 + y_Q^2 + z_Q^2 = 0,
\end{multline}
that is, the first term in the right-hand side of expansion~(\ref{eq:2:delta-s-via-A}) vanishes. Hence, due to Eq.~(\ref{eq:2:delta-s-via-A}), the interval $\delta s^2_{PQ}$ is proportional to the curvature tensor $R_{iklm}(O)$, up to the residual term $o(\epsilon^4)$. This fact can be expressed as follows:
\begin{equation}
\label{eq:2:PQ-interval-1}
\delta s^2_{PQ} = R_{iklm}(O) \, M_{PQ}^{iklm} + o(\epsilon^4),
\end{equation}
where tensor $M_{PQ}^{iklm}$ is the coefficient at $R_{iklm}(O)$ in Eq.~(\ref{eq:2:delta-s-via-A}):
\begin{equation}
\label{eq:2:M-PQ-def}
M_{PQ}^{iklm} = - \frac13 \, r_P^i \, (r_Q^k - r_P^k)\, r_P^l (r_Q^m - r_P^m) .
\end{equation}

Let us consider the following combination $\Sigma$ of sixteen intervals $\delta s^2_{PQ}$, where $P\in\{A,B,C,D\}$, and $Q\in\{E,F,G,H\}$:
\begin{eqnarray}
\Sigma &=& \delta s^2_{AE} + \delta s^2_{AF} - \delta s^2_{AG} - \delta s^2_{AH} \nonumber \\
       &+& \delta s^2_{BE} + \delta s^2_{BF} - \delta s^2_{BG} - \delta s^2_{BH} \nonumber \\
       &-& \delta s^2_{CE} - \delta s^2_{CF} + \delta s^2_{CG} + \delta s^2_{CH} \nonumber \\
       &-& \delta s^2_{DE} - \delta s^2_{DF} + \delta s^2_{DG} + \delta s^2_{DH}\, . \label{eq:2:Sigma-def}
\end{eqnarray}
The choice of signs in this definition enables a remarkable property of stability of $\Sigma$ under small variations of points $A$ -- $H$. We will establish this property in subsection~\ref{sec:stability}. 

It is evident from Eqs.~(\ref{eq:2:PQ-interval-1}) and~(\ref{eq:2:Sigma-def}) that quantity $\Sigma$ is also proportional to $R_{iklm}(O)$, up to the residual term:
\begin{equation}
\label{eq:2:Sigma-via-A}
\Sigma = R_{iklm}(O) \, M^{iklm} + o(\epsilon^4),
\end{equation}
where $M^{iklm}$ is a combination of tensors $M_{PQ}^{iklm}\,$:
\begin{eqnarray}
M &=& M_{AE} + M_{AF} - M_{AG} - M_{AH} \nonumber \\
  &+& M_{BE} + M_{BF} - M_{BG} - M_{BH} \nonumber \\
  &-& M_{CE} - M_{CF} + M_{CG} + M_{CH} \nonumber \\
  &-& M_{DE} - M_{DF} + M_{DG} + M_{DH} \label{eq:2:M-def}
\end{eqnarray}
(we have omitted tensor superscripts ``$iklm$'' in Eq.~(\ref{eq:2:M-def}) for brevity). 

Tensor $M^{iklm}$ can be calculated straightforwardly, by substituting of equations~(\ref{eq:2:A-def})--(\ref{eq:2:H-def}) and~(\ref{eq:2:M-PQ-def}) into Eq.~(\ref{eq:2:M-def}). Such a calculation however demands evaluation of the right-hand side of Eq.~(\ref{eq:2:M-PQ-def}) for each $P$, $Q$, $i$, $k$, $l$ and $m$, i.~e. $4 \times 4 \times 4 \times 4 \times 4 \times 4 = 4096$ times. We have done this with the aid of a computer (the Matlab code is presented in Appendix~\ref{app:code}), and obtained the following results:
\begin{eqnarray}
M^{tzxz} = M^{xztz} &=& \frac{8\sqrt2}{3} \epsilon^4 , \label{eq:2:M-result-1} \\
M^{tyxy} = M^{xyty} &=& - \frac{8\sqrt2}{3} \epsilon^4 , \label{eq:2:M-result-2} 
\end{eqnarray}
and all other components of tensor $M^{iklm}$ are equal to zero. 
Substituting these expressions for the components of $M^{iklm}$ into Eq.~(\ref{eq:2:Sigma-via-A}), one can find that
\begin{multline}
\label{eq:2:Sigma-via-A-2}
\Sigma = \frac{8\sqrt2}{3} \epsilon^4 \Big[ R_{tzxz}(O) + R_{xztz}(O) \\
- R_{tyxy}(O) - R_{xyty}(O) \Big] + o(\epsilon^4),
\end{multline}
that can be simplified using the permutation symmetry $R_{iklm} = R_{lmik}$:
\begin{equation}
\label{eq:2:Sigma-via-A-3}
\Sigma = 
\frac{16\sqrt2}{3} \epsilon^4 \left[ R_{tzxz}(O) - R_{tyxy}(O) \right]
+ o(\epsilon^4).
\end{equation}
It is clearly seen from Eq.~(\ref{eq:2:C-tyxy-via-R}) that the expression in square brackets is equal to $-2C_{tyxy}(O)$. Hence, the linear combination $\Sigma$ of intervals is expressed through the Weyl tensor component $C_{tyxy}$:
\begin{equation}
\label{eq:2:Sigma-via-C}
\Sigma = 
-\frac{32\sqrt2}{3} \, \epsilon^4 \, C_{tyxy}(O) + o(\epsilon^4) .
\end{equation}

\subsection{Stability of combination $\Sigma$ under small shifts of~points $A$ -- $H$}
\label{sec:stability}

The aim of this subsection is to find out how quantity $\Sigma$ changes under small variations of points $A, \ldots, H$. Let us choose eight points $A', \ldots, H'$ located near points $A, \ldots, H$ so that the difference between coordinates of the corresponding points is small with respect to $\epsilon^2$:
\begin{eqnarray}
|t_{A'}-t_A|=o(\epsilon^2), \quad 
|x_{A'}-x_A|=o(\epsilon^2), \nonumber\\ 
|y_{A'}-y_A|=o(\epsilon^2), \quad 
|z_{A'}-z_A|=o(\epsilon^2), \label{eq:2:A-Aprime-difference}
\end{eqnarray}
and so on. Then, we replace points $A, \ldots, H$ with their corresponding points $A', \ldots, H'$ in definition~(\ref{eq:2:Sigma-def}) of quantity $\Sigma$, and denote the result of this replacement as~$\Sigma'$: 
\begin{eqnarray}
\Sigma' &=& \delta s^2_{A'E'} + \delta s^2_{A'F'} - \delta s^2_{A'G'} - \delta s^2_{A'H'} \nonumber \\
        &+& \delta s^2_{B'E'} + \delta s^2_{B'F'} - \delta s^2_{B'G'} - \delta s^2_{B'H'} \nonumber \\
        &-& \delta s^2_{C'E'} - \delta s^2_{C'F'} + \delta s^2_{C'G'} + \delta s^2_{C'H'} \nonumber \\
        &-& \delta s^2_{D'E'} - \delta s^2_{D'F'} + \delta s^2_{D'G'} + \delta s^2_{D'H'}\, . \label{eq:2:Sigma-prime-def}
\end{eqnarray}
In this subsection, we will show that the difference between $\Sigma$ and $\Sigma'$ is small in comparison with $\epsilon^4$.

Let $P$ be any of points $\{A,B,C,D\}$, and $Q$ be any of points $\{E,F,G,H\}$. Let us find the difference between intervals $\delta s^2_{P'Q'}$ and $\delta s^2_{PQ}$ with the aid of expansion~(\ref{eq:2:delta-s-via-A}). First, we note that the second term in the right-hand side (rhs) of Eq.~(\ref{eq:2:delta-s-via-A}) is proportional to $\epsilon^4$. Therefore a change of this term, due to a small shift of points $P$ and $Q$, must be small with respect to $\epsilon^4$. 
Then, let us consider the first term in the rhs of Eq.~(\ref{eq:2:delta-s-via-A}). Essentially, this term is a Minkowski-space scalar product $(\r_Q-\r_P)\cdot(\r_Q-\r_P) \equiv (\r_Q-\r_P)^2$. Replacing $P$ and $Q$ with $P'$ and $Q'$ changes this term by amount of
\begin{multline} \label{eq:2:first-term-change}
(\r_{Q'}-\r_{P'})^2 - (\r_Q-\r_P)^2 \\
= 2(\r_Q-\r_P)\cdot(\overrightarrow{\Delta r}_Q-\overrightarrow{\Delta r}_P) 
+ (\overrightarrow{\Delta r}_Q-\overrightarrow{\Delta r}_P)^2 ,
\end{multline}
where
\begin{equation} \label{eq:2:Delta-r-def}
\overrightarrow{\Delta r}_P = \r_{P'} - \r_P \,,  \qquad
\overrightarrow{\Delta r}_Q = \r_{Q'} - \r_Q \,.
\end{equation}
According to our choice of points $P'$ and $Q'$, each component of vectors $\overrightarrow{\Delta r}_P$ and $\overrightarrow{\Delta r}_Q$ is small compared to $\epsilon^2$, see Eq.~(\ref{eq:2:A-Aprime-difference}). For this reason, the last term in Eq.~(\ref{eq:2:first-term-change}) is as small as $o(\epsilon^4)$. Therefore, the difference between intervals $\delta s^2_{P'Q'}$ and $\delta s^2_{PQ}$ is determined only by the first term in the rhs of Eq.~(\ref{eq:2:first-term-change}), up to $o(\epsilon^4)$:
\begin{equation} \label{eq:2:interval-change}
\delta s^2_{P'Q'} - \delta s^2_{PQ} 
= 2(\r_Q-\r_P)\cdot(\overrightarrow{\Delta r}_Q-\overrightarrow{\Delta r}_P) 
+ o(\epsilon^4).
\end{equation}

The difference between quantities $\Sigma'$, Eq.~(\ref{eq:2:Sigma-prime-def}), and $\Sigma$, Eq.~(\ref{eq:2:Sigma-def}), is a combination of expressions~(\ref{eq:2:interval-change}), taken with appropriate signs. After a simple algebra, one can represent this difference as follows:
\begin{multline} \label{eq:2:sigma-change-2}
\Sigma' - \Sigma = \\
- 2(\r_A+\r_B-\r_C-\r_D)\cdot
(\overrightarrow{\Delta r}_E+\overrightarrow{\Delta r}_F-\overrightarrow{\Delta r}_G-\overrightarrow{\Delta r}_H) \\
- 2(\r_E+\r_F-\r_G-\r_H)\cdot
(\overrightarrow{\Delta r}_A+\overrightarrow{\Delta r}_B-\overrightarrow{\Delta r}_C-\overrightarrow{\Delta r}_D) \\
+ o(\epsilon^4).
\end{multline}

Finally, looking at definitions of points $A$ -- $H$, Eqs.~(\ref{eq:2:A-def}) -- (\ref{eq:2:H-def}), one can see that
\begin{equation} \label{eq:2:r-opposite}
\r_A=-\r_B, \quad 
\r_C=-\r_D, \quad 
\r_E=-\r_F, \quad 
\r_G=-\r_H. 
\end{equation}
Substitution of these relations into Eq.~(\ref{eq:2:sigma-change-2}) gives rise to a simple result:
\begin{equation} \label{eq:2:sigma-change}
\Sigma' = \Sigma + o(\epsilon^4).
\end{equation}
Remarkably, all terms in the difference $\Sigma' - \Sigma$ cancel each other, except for the residual term $o(\epsilon^4)$. This cancellation became possible as a result of the special choice of signs in the definition of quantity $\Sigma$.

Hence, a shift of points $A$ -- $H$ by distances $o(\epsilon^2)$ results in as small change of quantity $\Sigma$ as $o(\epsilon^4)$.

\subsection{Making fifteen intervals equal to zero}
\label{sec:15zeros}

Generally, all sixteen intervals $\delta s^2_{AE}, \ldots, \delta s^2_{DH}$, that contribute to $\Sigma$, can be different from zero. But, as we will show in this subsection, it is possible to set \emph{fifteen} of them to zero by small shifts of points $A$ -- $H$. This will be done in three steps described below. Shifted points are denoted as $A'$ -- $H'$.

{\bf Step 1.} We do not shift points $A$, $B$, and $C$. That is, $A' = A$, $B' = B$, and $C' = C$.

{\bf Step 2.} For each of points $E$, $F$, $G$ and $H$, we leave one of coordinates unchanged, namely $z_E$, $z_F$, $y_G$ and $y_H$. Then we adjust other three coordinates such that three intervals between the given point and points $A'$, $B'$ and $C'$ vanish. Therefore we obtain shifted points $E'$, $F'$, $G'$ and $H'$ that obey relations
\begin{align}
\delta s^2_{A'E'} = 
\delta s^2_{A'F'} = 
\delta s^2_{A'G'} = 
\delta s^2_{A'H'} &= \nonumber\\
\delta s^2_{B'E'} = 
\delta s^2_{B'F'} = 
\delta s^2_{B'G'} = 
\delta s^2_{B'H'} &= \nonumber\\
\delta s^2_{C'E'} = 
\delta s^2_{C'F'} = 
\delta s^2_{C'G'} = 
\delta s^2_{C'H'} &= 0. \label{eq:2:12zeros}
\end{align}

{\bf Step 3.} For point $D$, we fix coordinate $t_D$, and adjust other three coordinates such that three intervals between this point and points $E'$, $F'$ and $G'$ vanish:
\begin{equation}
\delta s^2_{D'E'} = 
\delta s^2_{D'F'} = 
\delta s^2_{D'G'} = 0, \label{eq:2:3zeros}
\end{equation}
where $D'$ is the shifted point $D$.

As shown in Appendix~\ref{app:15zeros}, differences between initial points $A$ -- $H$ and their corresponding shifted points $A'$ -- $H'$ do not exceed ${\cal O}(\epsilon^3)$ along each coordinate. 
This allows us to apply the results of Subsection~\ref{sec:stability} to this set of points.

\subsection{Completion of the proof}
\label{sec:completion}

Let us recall the results obtained above. We consider a curved spacetime with a nonzero Weyl tensor at some point $O$. 
In Subsection~\ref{sec:frame}, we have chosen a local inertial (at point $O$) reference frame, in which component $C_{tyxy}$ of the Weyl tensor is nonzero at $O$. In Subsection~\ref{sec:intervals}, we have introduced such a definition of interval $\delta s^2_{PQ}$ between two points $P$ and $Q$ in a curved spacetime, that $\delta s^2_{PQ} = 0$ if and only if relation $P\lightcone Q$ is fulfilled. In Subsection~\ref{sec:combination}, we have defined eight points $A$ -- $H$ within the $\epsilon$-neighborhood of point $O$. Then we have found a linear combination $\Sigma$ of intervals between these points, which is proportional to $C_{tyxy}$ up to terms $o(\epsilon^4)$. In Subsection~\ref{sec:stability}, we have shown that this combination $\Sigma$ remains unchanged under small ($o(\epsilon^2)$) variations of coordinates of points $A$ -- $H$, up to $o(\epsilon^4)$. Finally, we have found in Subsection~\ref{sec:15zeros} such a variation of coordinates that enable to set fifteen intervals $\delta s^2_{A'E'}, \ldots, \delta s^2_{D'G'}$ to zero.

Collected together, these results consist a proof of Theorem~\ref{th:Weyl}. To see this, let us choose a small spatial scale~$\epsilon$, then define eight points $A$ -- $H$ in the $\epsilon$-neighborhood around point $O$ according to Subsection~\ref{sec:combination}, and find shifted points $A'$ -- $H'$ by the procedure described in Subsection~\ref{sec:15zeros}. As a result, all intervals, that define quantity $\Sigma'$ in Eq.~(\ref{eq:2:Sigma-prime-def}), vanish, except for $\delta s^2_{D'H'}$. Therefore
\begin{equation} \label{eq:2:Sigma-prime-one-interval}
\Sigma' = \delta s^2_{D'H'} \,.
\end{equation}
Taking into account stability of $\Sigma$ under small shifts, Eq.~(\ref{eq:2:sigma-change}), one can see that
\begin{equation} \label{eq:2:Sigma-one-interval}
\delta s^2_{D'H'} = \Sigma + o(\epsilon^4) .
\end{equation}
Then, expressing quantity $\Sigma$ through the Weyl tensor according to Eq.~(\ref{eq:2:Sigma-via-C}), one obtains
\begin{equation} \label{eq:2:one-interval-via-C}
\delta s^2_{D'H'} = 
-\frac{32\sqrt2}{3} \, \epsilon^4 \, C_{tyxy}(O) + o(\epsilon^4) .
\end{equation}
In the chosen reference frame, $C_{tyxy}(O) \neq 0$, see Subsection~\ref{sec:frame}. When spatial scale $\epsilon$ is sufficiently small, the term containing $\epsilon^4 \, C_{tyxy}(O)$ exceeds (in absolute value) the residual term $o(\epsilon^4)$ in Eq.~(\ref{eq:2:one-interval-via-C}). Therefore, for such value of $\epsilon$,
\begin{equation} \label{eq:2:one-interval-neq-0}
\delta s^2_{D'H'} \neq 0 .
\end{equation}
The set of points $A'$ -- $H'$ obeys Eqs.~(\ref{eq:2:12zeros}) and~(\ref{eq:2:3zeros}). According to Subsection~\ref{sec:intervals}, this means that fifteen relations
\begin{align} 
A'\lightcone E',&\;\; A'\lightcone F',\;\; A'\lightcone G',\;\; A'\lightcone H', \nonumber\\
B'\lightcone E',&\;\; B'\lightcone F',\;\; B'\lightcone G',\;\; B'\lightcone H', \nonumber\\
C'\lightcone E',&\;\; C'\lightcone F',\;\; C'\lightcone G',\;\; C'\lightcone H', \nonumber\\
D'\lightcone E',&\;\; D'\lightcone F',\;\; D'\lightcone G' \label{eq:2:15relations}
\end{align}
are fulfilled. 

If the spacetime were well-stitched, relation $D'\lightcone H'$ would follow from relations~(\ref{eq:2:15relations}). But Eq.~(\ref{eq:2:one-interval-neq-0}) means that relation $D'\lightcone H'$ is not fulfilled. Thus, points $A'$ -- $H'$ violate well-stitchedness.

In conclusion, provided that the Weyl tensor differs from zero at some point $O$, there are eight points $A'$ -- $H'$ that violate well-stitchedness. Q.E.D.

\subsection{Estimate of the time delay}
\label{sec:delta-t}

As a by-product of the proof presented above, we have obtained the value of $\delta s^2_{D'H'}$, which is a measure of violation of relation $D'\lightcone H'$. Inequality $\delta s^2_{D'H'} \neq 0$ means that a light signal emitted at event $D'$ does not pass through event $H'$. We denote as $\delta t_{D'H'}$ the time delay of this signal, i.~e. the difference between the moment of time, when this signal reaches point $H'$, and the moment of time of event $H'$ itself. We are going to find the value of this time delay.

\begin{figure}
\includegraphics[width=4cm]{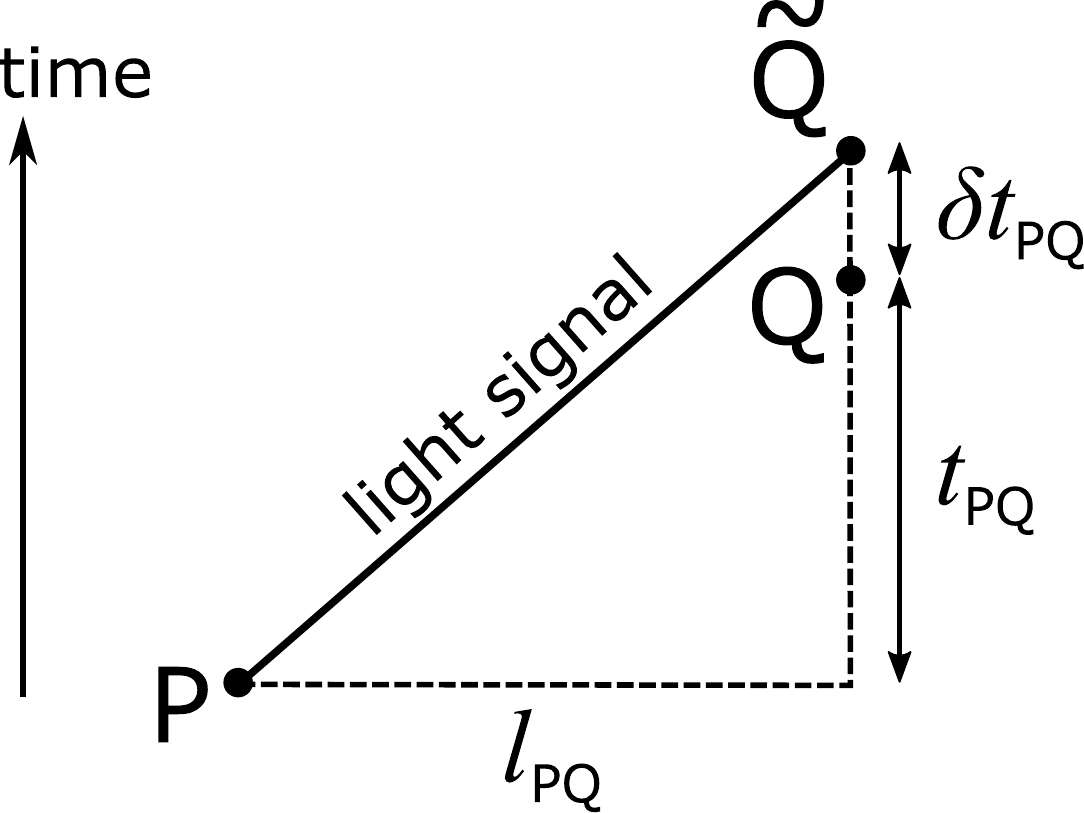}
\caption{Illustration of time delay $\delta t_{PQ}$. A light signal emitted at event $P$ passes in a vicinity of event $Q$. The event when this signal comes to the point of space of event $Q$ is denoted as $\widetilde Q$. Time delay $\delta t_{PQ}$ is the difference in time between events $Q$ and $\widetilde Q$.}
\label{fig:delta_t}
\end{figure}

For the sake of generality, let us consider two events $P$ and $Q$, and denote as $\widetilde Q$ the event, in which a light signal from $P$ reaches the spatial point of event $Q$ (see Fig.~\ref{fig:delta_t}). One can make the reference frame locally-inertial (i. e. set Christoffel symbols to zero) along line $P\widetilde Q$ by a small distortion of coordinates~\cite[\S85]{LL2}. In such a frame, the light travels from $P$ to $\widetilde Q$ with the constant velocity $c$, whence
\begin{equation} \label{eq:2:l-PQ}
\ell_{PQ} = c(t_{PQ} + \delta t_{PQ}) ,
\end{equation}
where $\ell_{PQ}$ and $t_{PQ}$ are differences between $P$ and $Q$ in space and in time, correspondingly. Interval $\delta s^2_{PQ}$ is expressed in this frame just as in special relativity:
\begin{equation} \label{eq:2:interval-PQ}
\delta s^2_{PQ} = \ell_{PQ}^2 - c^2 t_{PQ}^2 .
\end{equation}
Substituting Eq.~(\ref{eq:2:l-PQ}) into Eq.~(\ref{eq:2:interval-PQ}) and neglecting the higher-order infinitesimal $\delta t_{PQ}^2$, one can express time delay $\delta t_{PQ}$ as follows (cf. Ref.~\cite{Teyssandier2008}):
\begin{equation} \label{eq:2:delta-t-PQ}
\delta t_{PQ} \approx \frac{\delta s^2_{PQ}}{2c^2t_{PQ}} \equiv \frac{\delta s^2_{PQ}}{2c^2(t_Q - t_P)} \, .
\end{equation}

Taking $\delta s^2_{D'H'}$ from Eq.~(\ref{eq:2:one-interval-via-C}) and $t_{H'} - t_{D'} \approx t_H - t_D = \sqrt2 \epsilon/c$ from Eq.~(\ref{eq:2:AH-def}), one can obtain with the help of Eq.~(\ref{eq:2:delta-t-PQ}) that
\begin{equation} \label{eq:2:delta-t-via-C}
\delta t_{D'H'} \approx 
-\frac{16 \epsilon^3 \, C_{tyxy}(O)}{3c} \, .
\end{equation}
This is the time delay, detection of which reveals the spacetime curvature.

\newpage
\appendix


\section{Proof of Lemma~\ref{lemma:scalar-product}}
\label{app:proof-scalar-product}

{\bf Lemma~\ref{lemma:scalar-product}.} 
For any four points $P$, $Q$, $R$, $S$, if relations $P\lightcone R$, $P\lightcone S$, $Q\lightcone R$ and $Q\lightcone S$ are fulfilled, then vector $\overrightarrow{PQ}$ is perpendicular to vector $\overrightarrow{RS}$ (that is, $\overrightarrow{PQ} \cdot \overrightarrow{RS} = 0$).

{\bf Proof.} Let $\r_P$, $\r_Q$, $\r_R$ and $\r_S$ be radius-vectors of points $P$, $Q$, $R$ and $S$, correspondingly. Then, $\overrightarrow{PQ} = \r_Q-\r_P$, and $\overrightarrow{RS} = \r_S-\r_R$. Relations $P\lightcone R$, $P\lightcone S$, $Q\lightcone R$ and $Q\lightcone S$ mean that
\begin{gather}
\label{eq:app:PR}
(\r_P-\r_R) \cdot (\r_P-\r_R) = 0, \\
\label{eq:app:PS}
(\r_P-\r_S) \cdot (\r_P-\r_S) = 0, \\
\label{eq:app:QR}
(\r_Q-\r_R) \cdot (\r_Q-\r_R) = 0, \\
\label{eq:app:QS}
(\r_Q-\r_S) \cdot (\r_Q-\r_S) = 0.
\end{gather}
Summing up Eqs.~(\ref{eq:app:PS}) and~(\ref{eq:app:QR}), and subtracting Eqs.~(\ref{eq:app:PR}) and~(\ref{eq:app:QS}) from them, and expanding all brackets, one can get
\begin{equation}
\label{eq:app:sum1}
2\r_P\cdot\r_R - 2\r_P\cdot\r_S - 2\r_Q\cdot\r_R + 2\r_Q\cdot\r_S = 0,
\end{equation}
that is, 
\begin{equation}
\label{eq:app:sum2}
2(\r_Q-\r_P)\cdot(\r_S-\r_R) \equiv 2 \overrightarrow{PQ} \cdot \overrightarrow{RS} = 0,
\end{equation}
Q.E.D.


\section{Proof of Lemma~\ref{lemma:one-line}}
\label{app:proof-one-line}

{\bf Lemma~\ref{lemma:one-line}.} 
For any three different points $P$, $Q$, $R$, if relations $P\lightcone Q$, $P\lightcone R$ and $Q\lightcone R$ are fulfilled, then these points lie on one line.

{\bf Proof.} Let us choose the coordinate system with the origin at point $P$. It follows from relation $P\lightcone Q$ that point $Q$ lies on the light cone with the vertex at $P$:
\begin{equation}
\label{eq:app:Q-light-cone}
-t_Q^2 + x_Q^2 + y_Q^2 + z_Q^2 = 0.
\end{equation}
Let us turn the axes $x$, $y$, $z$ such that point $Q$ lie in plane $tx$. Therefore, according to Eq.~(\ref{eq:app:Q-light-cone}),
\begin{equation}
\label{eq:app:Q-eq}
x_Q = t_Q \neq 0, \quad y_Q = z_Q = 0.
\end{equation}
[Another option is $x_Q = -t_Q$, but in this case we will flip the direction of axis $x$ and return to Eq.~(\ref{eq:app:Q-eq}).] 
Due to relations $P\lightcone R$ and $Q\lightcone R$, point $R$ must lie on both light cones---with vertices at $P$ and $Q$:
\begin{gather}
\label{eq:app:R-light-cone1}
-t_R^2 + x_R^2 + y_R^2 + z_R^2 = 0, \\
\label{eq:app:R-light-cone2}
-(t_R-t_Q)^2 + (x_R-t_Q)^2 + y_R^2 + z_R^2 = 0.
\end{gather}
Subtracting Eq.~(\ref{eq:app:R-light-cone1}) from Eq.~(\ref{eq:app:R-light-cone2}) yields
\begin{equation}
\label{eq:app:R-eq1}
2t_Q (t_R - x_R) = 0,
\end{equation}
i.~e. $x_R = t_R$. Substituting $x_R = t_R$ into Eq.~(\ref{eq:app:R-light-cone1}), one can find that $y_R = z_R = 0$.

Hence, all three points $P$, $Q$, $R$ belong to the same (light-like) line, which equation is
\begin{equation}
\label{eq:PQR-line}
x = t, \quad y = 0, \quad z = 0.
\end{equation}
Q.E.D.


\section{Proof of Lemma~\ref{lemma:light-like}}
\label{app:proof-light-like}

{\bf Lemma~\ref{lemma:light-like}.} 
For any four points $P$, $Q$, $R$, $S$, if relations $P\lightcone S$, $Q\lightcone S$ and $R\lightcone S$ are fulfilled, and points $P$, $Q$, $R$ lie on one line and are different from each other, then this line is light-like (that is, $P\lightcone Q$, $P\lightcone R$ and $Q\lightcone R$).

{\bf Proof.} The line, that joins points $P$, $Q$ and $R$, may be time-like, space-like or light-like. Let us suppose first that this line is time-like. In such a case, this line can be chosen as $t$-axis. With this choice of coordinates, relations $P\lightcone S$ and $Q\lightcone S$ obtain the following form:
\begin{gather}
\label{eq:app:PS-cone}
x_S^2 + y_S^2 + z_S^2 = (t_S-t_P)^2, \\
\label{eq:app:QS-cone}
x_S^2 + y_S^2 + z_S^2 = (t_S-t_Q)^2.
\end{gather}
Therefore $|t_S-t_P| = |t_S-t_Q|$. Taking into account that $t_P \neq t_Q$ (since points $P$ and $Q$ are different), one can resolve the latter equality as
\begin{equation}
\label{eq:t-S-PQ}
t_S = \frac{t_P+t_Q}{2}.
\end{equation}
Similarly, one can deduce from relations $P\lightcone S$ and $R\lightcone S$ that
\begin{equation}
\label{eq:t-S-PR}
t_S = \frac{t_P+t_R}{2}.
\end{equation}
It is clearly seen from Eqs.~(\ref{eq:t-S-PQ}) and~(\ref{eq:t-S-PR}) that $t_Q = t_R$. But this contradicts to the proposition that points $Q$ and $R$ are different. Consequently, the line that passes through $P$, $Q$ and $R$ cannot be time-like.

Next, we suppose that this line is space-like. Let us choose this line as $x$-axis. Then, almost the same reasoning, as in the case of a time-like line, leads from relations $P\lightcone S$, $Q\lightcone S$ and $R\lightcone S$ to equations 
\begin{gather}
\label{eq:x-S-PQ}
x_S = \frac{x_P+x_Q}{2}, \\
\label{eq:x-S-PR}
x_S = \frac{x_P+x_R}{2},
\end{gather}
and to conclusion that $x_Q = x_R$. And again, this contradicts to the proposition that $Q$ and $R$ are different. This means that the line that connects points $P$, $Q$, $R$ is not space-like.

Hence, this line is neither time-like nor space-like. It is therefore light-like, Q.E.D.


\section{The case when neither plane $ABCD$ nor plane $EFGH$ contains a time-like vector}
\label{app:option-c}

Let $\alpha$ and $\beta$ be 2-dimensional planes in 4-dimensional Minkowski space, and any vector in $\alpha$ is orthogonal to any vector in $\beta$. (In the main text, $\alpha$ is plane $ABCD$, and $\beta$ is plane $EFGH$.) We suppose also that neither $\alpha$ nor $\beta$ contains a time-like vector. The aim of this Appendix is to show that, in a properly chosen coordinate frame, planes $\alpha$ and $\beta$ are defined by Eqs.~(\ref{eq:1:ABCD-plane-option-c}) and~(\ref{eq:1:EFGH-plane-option-c}), respectively.

Let us choose two non-collinear, mutually perpendicular vectors $\vec a$ and $\vec b$ in plane $\alpha$. Each of these vectors can be either space-like, or light-like. (Time-like vectors are forbidden by supposition.) There are three cases: (i) both vectors are space-like, (ii) both are light-like, and (iii) one vector is space-like and the other one is light-like.

In case (i), one can choose directions of coordinate axis $y$ and $z$ along vectors $\vec a$ and $\vec b$, correspondingly. With this choice, plane $\alpha$ is parallel to $yz$-plane, and consequently any vector of plane $\beta$ is perpendicular to $yz$-plane, i.~e. parallel to $tx$-plane. Therefore plane $\beta$ is parallel to plane $tx$. But it contradicts to the supposition that plane $\beta$ does not contain a time-like vector. 

Case (ii) contradicts to the choice of vectors $\vec a$ and $\vec b$ as non-collinear and mutually perpendicular. Indeed, two light-like vectors in the Minkowski space can be perpendicular to each other only if they are collinear.

Therefore cases (i) and (ii) lead to contradictions, and thus one of the vectors is light-like and the other is space-like. Let $\vec a$ be a light-like vector, and $\vec b$ be a space-like one. One can choose the coordinate axes such that vector $\vec a$ lie in $tx$-plane, and its $t$- and $x$-components be the same:
\begin{equation}
\label{eq:app:a-in-tx}
x_a = t_a \neq 0, \quad y_a = z_a = 0.
\end{equation}
It follows from Eq.~(\ref{eq:app:a-in-tx}) and from orthogonality of vectors $\vec a$ and $\vec b$ that $x_b = t_b$. By rotation of $yz$-part of the coordinate system, one can put $z$-component of vector $\vec b$ to zero, while components of vector $\vec a$ remain unchanged. After this, components of vector $\vec b$ obeys the following conditions:
\begin{equation}
\label{eq:app:b-in-txy}
x_b = t_b, \quad z_b = 0, \quad y_b \neq 0
\end{equation}
(the latter inequality follows from non-collinearity of vectors $\vec a$ and $\vec b$).
Equations~(\ref{eq:app:a-in-tx}) and~(\ref{eq:app:b-in-txy}) show that vectors $\vec a$ and $\vec b$ are linear combinations of the following two vectors:
\begin{equation}
\label{eq:app:cd}
\vec c = (1,1,0,0) \text{ and } \vec d = (0,0,1,0),
\end{equation}
where the components are listed in order $(t,x,y,z)$. 
Therefore, radius-vector $\r_\alpha$ of an arbitrary point on plane $\alpha$ can be expressed as
\begin{equation}
\label{eq:app:r-alpha}
\r_\alpha = \r_{\alpha0} + \lambda\vec c + \mu\vec d,
\end{equation}
where $\lambda$ and $\mu$ are real parameters that vary from point to point, and $\r_{\alpha0}$ is a radius-vector of some fixed point on $\alpha$. 
Similarly, radius-vector $\r_\beta$ of an arbitrary point on plane $\beta$ depends on two real parameters $\lambda'$ and $\mu'$ as follows:
\begin{equation}
\label{eq:app:r-beta}
\r_\beta = \r_{\beta0} + \lambda'\vec e + \mu'\vec f,
\end{equation}
where $\r_{\beta0}$ is a radius-vector of some fixed point on plane $\beta$, and each of two non-collinear vectors $\vec e$ and $\vec f$ is perpendicular to both $\vec c$ and $\vec d$. It is convenient to choose vectors $\vec e$ and $\vec f$ as
\begin{equation}
\label{eq:app:ef}
\vec e = (1,1,0,0) \text{ and } \vec f = (0,0,0,1).
\end{equation}
One can easily check that $\vec c \cdot \vec e = \vec c \cdot \vec f = \vec d \cdot \vec e = \vec d \cdot \vec f = 0$.

Now we will choose points $\r_{\alpha0}$ and $\r_{\beta0}$. First, we take randomly chosen points of planes $\alpha$ and $\beta$ as $\r_{\alpha0}$ and $\r_{\beta0}$. Then we shift point $\r_{\alpha0}$ along $y$-axis (i.~e. along $\vec d$), until its $y$-coordinate becomes equal to the $y$-coordinate of point $\r_{\beta0}$. Next, we shift point $\r_{\beta0}$ along $z$-axis (i.~e. along $\vec f$), until its $z$-coordinate becomes equal to that of point $\r_{\alpha0}$. And finally, we shift point $\r_{\beta0}$ along vector $\vec e$, until its $x$-coordinate becomes equal to that of point $\r_{\alpha0}$. In the end of these manipulations, points $\r_{\alpha0}$ and $\r_{\beta0}$ differ from each other only by their $t$-coordinates. Let us put the origin of coordinates to point $\r_{\alpha0}$. Therefore
\begin{equation}
\label{eq:app:r-alpha0-beta0}
\r_{\alpha0} = (0,0,0,0), \quad  \r_{\beta0} = (s,0,0,0),
\end{equation}
where $s$ is the difference between $t$-coordinates of these points.

One can see from Eqs.~(\ref{eq:app:cd}), (\ref{eq:app:r-alpha}) and~(\ref{eq:app:r-alpha0-beta0}) that coordinates $(t,x,y,z)$ of any point on plane $\alpha$ are
\begin{equation}
\label{eq:app:txyz-alpha}
t=\lambda, \quad x=\lambda, \quad y=\mu, \quad z=0,
\end{equation}
where $\lambda$ and $\mu$ are real parameters. This means that plane $\alpha$ is defined by Eq.~(\ref{eq:1:ABCD-plane-option-c}).

Similarly, Eqs.~(\ref{eq:app:r-beta}), (\ref{eq:app:ef}) and~(\ref{eq:app:r-alpha0-beta0}) express the coordinates of any point on plane $\beta$ through real parameters $\lambda'$ and $\mu'$:
\begin{equation}
\label{eq:app:txyz-beta}
t=\lambda'+s, \quad x=\lambda', \quad y=0, \quad z=\mu',
\end{equation}
that defines plane $\beta$ according to Eq.~(\ref{eq:1:EFGH-plane-option-c}).

Hence, equations~(\ref{eq:1:ABCD-plane-option-c}) and~(\ref{eq:1:EFGH-plane-option-c}) for planes $\alpha$ and $\beta$ are established.


\section{Possibility of choosing a reference frame, in which $C_{tyxy}\neq0$}
\label{app:C-inequality}

In this Appendix, we consider Weyl tensor $C_{iklm}$ at some given point $O$. We suppose that the reference frame is local inertial at this point, and therefore the metric tensor is equal to $\eta_{ik}$, i. e. to the metric tensor of special relativity.

All possible realizations of the Weyl tensor consist an irreducible representation $\cal R$ of the Lorentz group~\cite{Strichartz1988}. Suppose that there are such non-zero realizations of the Weyl tensor, in which its component $C_{tyxy}$ is equal to zero, and remains zero after any Lorentz transformation. A set of such realizations is also some representation ${\cal R}_0$ of the Lorentz group. By definition ${\cal R}_0 \subset \cal R$. Also ${\cal R}_0 \neq \cal R$, since a realization with $C_{tyxy} \neq 0$ belongs to $\cal R$ but not to ${\cal R}_0$. But this contradicts to irreducibility of $\cal R$. 

It is therefore proven by contradiction, that any non-zero Weyl tensor $C_{iklm}$ can be such Lorentz-transformed that inequality $C_{tyxy}\neq0$ will hold after the transformation.

\section{MATLAB code for calculation of tensor $M^{iklm}$}
\label{app:code}

\begin{verbatim}
components = 'txyz';

r_ABCD = [ sqrt(2),  1, 0, 0;
          -sqrt(2), -1, 0, 0;
           sqrt(2), -1, 0, 0;
          -sqrt(2),  1, 0, 0];

r_EFGH = [ 0, 0,  1,  0;
           0, 0, -1,  0;
           0, 0,  0,  1;
           0, 0,  0, -1];

sign_PQ = [ 1,  1, -1, -1;
            1,  1, -1, -1;
           -1, -1,  1,  1;
           -1, -1,  1,  1];

for i = 1:4
 for k = 1:4
  for l = 1:4
   for m = 1:4
    M_iklm = 0;
     for P = 1:4
      for Q = 1:4
       r_P = r_ABCD(P,:);
       r_Q = r_EFGH(Q,:);
       M_PQ_iklm = -1/3 * r_P(i) ...
        * (r_Q(k)-r_P(k)) * r_P(l) ...
        * (r_Q(m)-r_P(m));
       M_iklm = M_iklm ...
        + sign_PQ(P,Q) * M_PQ_iklm;
      end
     end
    if abs(M_iklm) > 1e-10
     disp(sprintf( 'M_%s = %f', ...
      components([i,k,l,m]), M_iklm ));
    end
   end
  end
 end
end
\end{verbatim}


\section{Estimation of distances between points $A$ -- $H$ and points $A'$ -- $H'$}
\label{app:15zeros}

In Subsection~\ref{sec:15zeros}, we suggested a method of finding such a set of ``shifted'' points $A'$ -- $H'$ that fifteen intervals listed in Eqs.~(\ref{eq:2:12zeros}) and~(\ref{eq:2:3zeros}) vanish. Here we discuss how close each ``shifted'' point is to its ``unshifted'' counterpart $A$ -- $H$.

For points $A'$, $B'$ and $C'$, this question is trivial because they coincide with $A$, $B$ and $C$, correspondingly.

Let us consider point $E'$. Its position is defined by a system of three equations
\begin{equation} \label{eq:app:AE-BE-CE-zero}
\delta s^2_{A'E'} = \delta s^2_{B'E'} = \delta s^2_{C'E'} = 0
\end{equation}
with three unknowns---namely, three coordinates $t_{E'}$, $x_{E'}$ and $y_{E'}$ of point $E'$. The fourth coordinate $z_{E'}$ is fixed:
\begin{equation} \label{eq:app:z-E-fixed}
z_{E'} = z_E .
\end{equation}
Equations~(\ref{eq:app:AE-BE-CE-zero}) are nonlinear, but can be approximately linearized when the deviation of point $E'$ from $E$ is small in comparison with the spatial scale $\epsilon$. Omitting the residual (nonlinear) term $o(\epsilon^4)$ in Eq.~(\ref{eq:2:interval-change}), and taking into account that $\r_{A'}-\r_A = \r_{B'}-\r_B = \r_{C'}-\r_C = 0$, one can get the following expressions for intervals $\delta s^2_{A'E'}$, $\delta s^2_{B'E'}$ and $\delta s^2_{C'E'}$:
\begin{eqnarray}
\delta s^2_{A'E'} &\approx& 2(\r_E-\r_A)\cdot(\r_{E'}-\r_E) + \delta s^2_{AE} \,, \label{eq:app:AE-approx} \\
\delta s^2_{B'E'} &\approx& 2(\r_E-\r_B)\cdot(\r_{E'}-\r_E) + \delta s^2_{BE} \,, \label{eq:app:BE-approx} \\
\delta s^2_{C'E'} &\approx& 2(\r_E-\r_C)\cdot(\r_{E'}-\r_E) + \delta s^2_{CE} \,. \label{eq:app:CE-approx}
\end{eqnarray}
Substituting Eqs.~(\ref{eq:app:z-E-fixed}) -- (\ref{eq:app:CE-approx}) into (\ref{eq:app:AE-BE-CE-zero}), we obtain a system of linearized equations. It is convenient to represent this system in a matrix form:
\begin{equation} \label{eq:app:AE-BE-CE-zero-via-matrix}
\hat A
\begin{pmatrix}
t_{E'} - t_E \\ x_{E'} - x_E \\ y_{E'} - y_E 
\end{pmatrix}
+
\begin{pmatrix}
\delta s^2_{AE} \\ \delta s^2_{BE} \\ \delta s^2_{CE} 
\end{pmatrix}
\approx
\begin{pmatrix}
0 \\ 0 \\ 0 
\end{pmatrix}
,
\end{equation}
where $\hat A$ is a $3\times3$ matrix of coefficients:
\begin{equation} \label{eq:app:A-matrix-def}
\hat A = 2
\begin{pmatrix}
t_A-t_E & x_E-x_A & y_E-y_A \\
t_B-t_E & x_E-x_B & y_E-y_B \\
t_C-t_E & x_E-x_C & y_E-y_C 
\end{pmatrix}
= 2\epsilon
\begin{pmatrix}
\;\;\sqrt2 &    -1 & 1 \\
   -\sqrt2 & \;\;1 & 1 \\
\;\;\sqrt2 & \;\;1 & 1 \\
\end{pmatrix}
\end{equation}
Matrix $\hat A$ is non-degenerate, and therefore system of equations~(\ref{eq:app:AE-BE-CE-zero-via-matrix}) is consistent. Its solution reads:
\begin{equation} \label{eq:app:AE-BE-CE-zero-solution}
\begin{pmatrix}
t_{E'} - t_E \\ x_{E'} - x_E \\ y_{E'} - y_E 
\end{pmatrix}
\approx -\hat A^{-1}
\begin{pmatrix}
\delta s^2_{AE} \\ \delta s^2_{BE} \\ \delta s^2_{CE} 
\end{pmatrix}
.
\end{equation}
The right-hand side of the latter equation is proportional to $\epsilon^3$. Indeed, coefficients of the inverse matrix $\hat A^{-1}$ are proportional to $\epsilon^{-1}$. Intervals $\delta s^2_{AE}$, $\delta s^2_{BE}$ and $\delta s^2_{CE}$ are proportional to $\epsilon^4$, that can be seen in Eq.~(\ref{eq:2:delta-s-via-A}). Hence,
\begin{eqnarray}
|t_{E'} - t_E| = {\cal O}(\epsilon^3), \quad
|x_{E'} - x_E| = {\cal O}(\epsilon^3), \nonumber\\
|y_{E'} - y_E| = {\cal O}(\epsilon^3), \quad 
|z_{E'} - z_E| = 0. \label{eq:app:E-shift-estim}
\end{eqnarray}

Similar considerations for point $F'$ lead to a linearized system of three equations with three unknowns $t_{F'}$, $x_{F'}$ and $y_{F'}$. This system can be written in the same form as Eq.~(\ref{eq:app:AE-BE-CE-zero-via-matrix}), where $E,E'$ are replaced with $F,F'$, and matrix $\hat A$ is slightly different:
\begin{equation} \label{eq:app:A-matrix-def-F}
\hat A = 2\epsilon
\begin{pmatrix}
\;\;\sqrt2 &    -1 & -1 \\
   -\sqrt2 & \;\;1 & -1 \\
\;\;\sqrt2 & \;\;1 & -1 \\
\end{pmatrix}
.
\end{equation}
As a result, one can get the same estimates for coordinate differences between points $F'$ and $F$ as in Eq.~(\ref{eq:app:E-shift-estim}).

For points $G'$ and $H'$, the analysis is exactly the same as for points $E'$ and $F'$, up to replacements $E \to G$, $E' \to G'$, $F \to H$, $F' \to H'$, and $y \leftrightarrow z$. It leads to the conclusion that points $E'$, $F'$, $G'$, $H'$ differ from their corresponding ``unshifted'' points $E$, $F$, $G$, $H$ by at most ${\cal O}(\epsilon^3)$ along each coordinate.

The last point to consider is $D'$. This point is defined by the system of three equations
\begin{equation} \label{eq:app:DE-DF-DG-zero}
\delta s^2_{D'E'} = \delta s^2_{D'F'} = \delta s^2_{D'G'} = 0
\end{equation}
with respect to three unknowns---coordinates $x_{D'}$, $y_{D'}$ and $z_{D'}$, whereas the fourth coordinate $t_{D'}$ is fixed: $t_{D'} = t_D$. One can express these intervals via Eq.~(\ref{eq:2:interval-change}), neglecting the nonlinear residual term $o(\epsilon^4)$:
\begin{eqnarray}
\delta s^2_{D'E'} \approx 2(\r_D-\r_E)\cdot(\r_{D'}-\r_D+\r_E-\r_{E'}) + \delta s^2_{DE} \,, \nonumber\\
\delta s^2_{D'F'} \approx 2(\r_D-\r_F)\cdot(\r_{D'}-\r_D+\r_F-\r_{F'}) + \delta s^2_{DF} \,, \nonumber\\
\delta s^2_{D'G'} \approx 2(\r_D-\r_G)\cdot(\r_{D'}-\r_D+\r_G-\r_{G'}) + \delta s^2_{DG} \,. \nonumber
\end{eqnarray}
These expressions provide a linearized form of the system of equations~(\ref{eq:app:DE-DF-DG-zero}):
\begin{equation} \label{eq:app:DE-DF-DG-zero-via-matrix}
\hat B
\begin{pmatrix}
x_{D'} - x_D \\ y_{D'} - y_D \\ z_{D'} - z_D 
\end{pmatrix}
+
\begin{pmatrix}
b_E \\ b_F \\ b_G 
\end{pmatrix}
\approx
\begin{pmatrix}
0 \\ 0 \\ 0 
\end{pmatrix}
,
\end{equation}
where matrix $\hat B$ is defined as follows:
\begin{equation} \label{eq:app:B-matrix-def}
\hat B = 2
\begin{pmatrix}
x_D-x_E & y_D-y_E & z_D-z_E \\
x_D-x_F & y_D-y_F & z_D-z_F \\
x_D-x_G & y_D-y_G & z_D-z_G
\end{pmatrix}
= 2\epsilon
\begin{pmatrix}
1 &    -1 & \;\;0 \\
1 & \;\;1 & \;\;0 \\
1 & \;\;0 &    -1 \\
\end{pmatrix}
\end{equation}
and $b_E$, $b_F$, $b_G$ denote the constant terms:
\begin{eqnarray}
b_E = 2(\r_D-\r_E)\cdot(\r_E-\r_{E'}) + \delta s^2_{DE} \,, \label{eq:app:b-E-def}\\
b_F = 2(\r_D-\r_F)\cdot(\r_F-\r_{F'}) + \delta s^2_{DF} \,, \label{eq:app:b-F-def}\\
b_G = 2(\r_D-\r_G)\cdot(\r_G-\r_{G'}) + \delta s^2_{DG} \,. \label{eq:app:b-G-def}
\end{eqnarray}
In the right-hand side of Eq.~(\ref{eq:app:b-E-def}), vector $(\r_D-\r_E)$ is proportional to $\epsilon$, see definitions~(\ref{eq:2:D-def}) and~(\ref{eq:2:E-def}); vector $(\r_E-\r_{E'})$ is proportional to $\epsilon^3$ due to Eq.~(\ref{eq:app:E-shift-estim}); and interval $\delta s^2_{DE}$ is proportional to $\epsilon^4$ according to Eq.~(\ref{eq:2:delta-s-via-A}). Therefore $b_E = {\cal O}(\epsilon^4)$, and the same estimates are true for quantities $b_F$ and $b_G$.

The solution of system~(\ref{eq:app:DE-DF-DG-zero-via-matrix}) is
\begin{equation} \label{eq:app:DE-DF-DG-zero-solution}
\begin{pmatrix}
x_{D'} - x_D \\ y_{D'} - y_D \\ z_{D'} - z_D 
\end{pmatrix}
\approx -\hat B^{-1}
\begin{pmatrix}
b_E \\ b_F \\ b_G 
\end{pmatrix}
.
\end{equation}
Here, matrix elements of $\hat B^{-1}$ are proportional to $\epsilon^{-1}$, and terms $b_E$, $b_F$, $b_G$ are proportional to $\epsilon^4$. Consequently the left-hand side of Eq.~(\ref{eq:app:DE-DF-DG-zero-solution}) is proportional to~$\epsilon^3$.

Hence, deviations of ``shifted'' points $A'$ -- $H'$ from ``unshifted'' ones $A$ -- $H$ do not exceed ${\cal O}(\epsilon^3)$.


\section{Calculation of the distance between geodesics in a complete quadrilateral}
\label{app:four-geodesics}

Let us consider a figure shown in Fig.~\ref{fig:quadrilateral_calc} (a complete quadrilateral) that consists of four geodesic lines: $OAB$, $OCD$, $AD$ and $BC$. 
If the spacetime were flat, geodesics $AD$ and $BC$ would cross each other at some point (event) $E_0$. In a curved spacetime, however, there may be a gap between lines $AD$ and $BC$. In this appendix, we address the following questions:

\begin{figure}
\includegraphics[width=5cm]{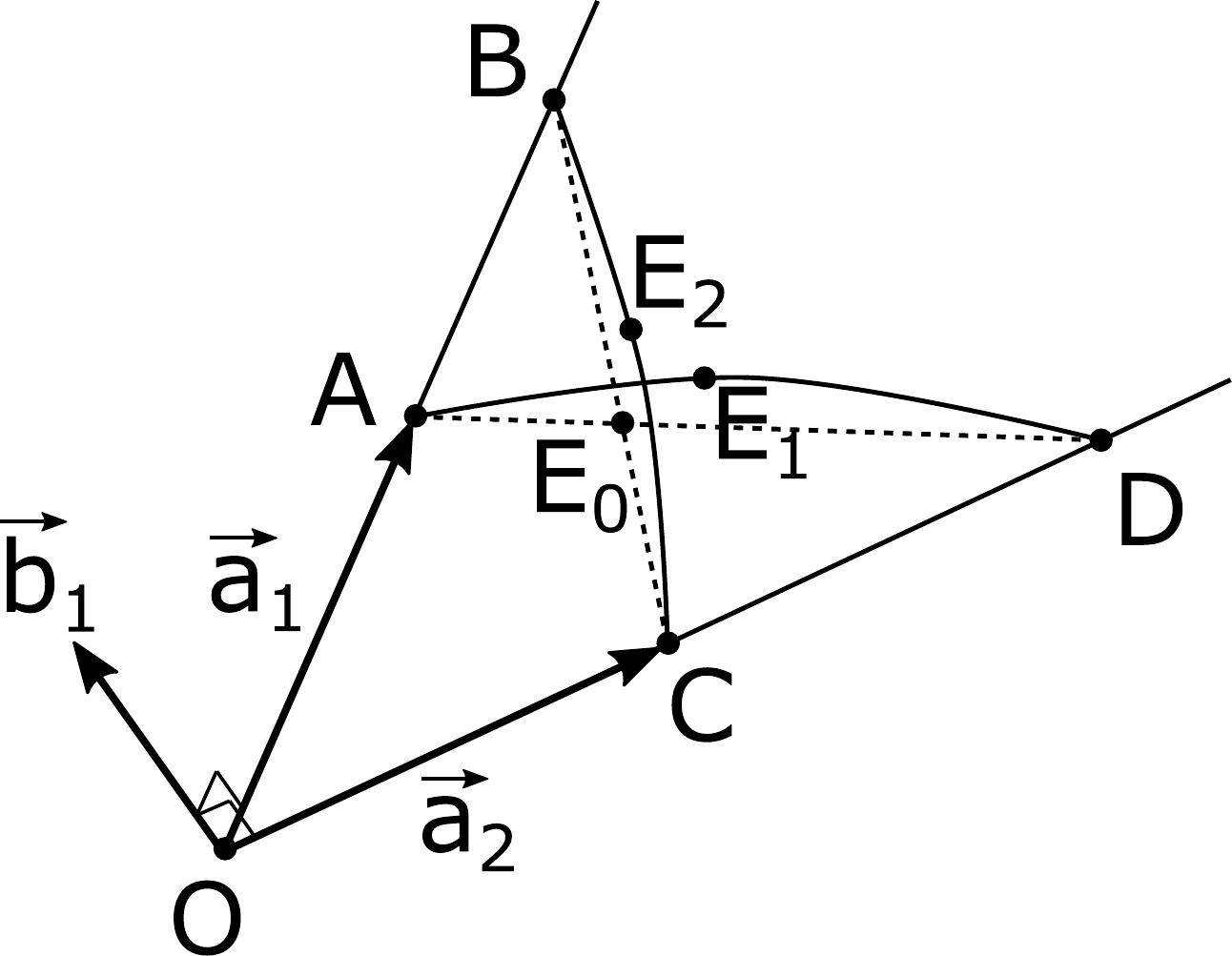}
\caption{A complete quadrilateral in a curved space. Solid lines $OAB$, $OCD$, $AE_1D$ and $BE_2C$ are geodesics. Dashed lines show that geodesics $AB$ and $CD$ would intersect in some point (event) $E_0$ if the spacetime were flat. Points $E_1$ and $E_2$ are chosen in a vicinity of point $E_0$. Unit vectors $\vec b_1$ and $\vec b_2$ (not shown) are perpendicular both to $\vec a_1$ and to $\vec a_2$.}
\label{fig:quadrilateral_calc}
\end{figure}

\textbullet~How to describe the size of the gap between lines $AD$ and $BC$?

\textbullet~How the size of this gap is related to the spacetime curvature?

\textbullet~To which extent the curvature tensor can be restored from the size of these gap (measured in different configurations of points $O$, $A$, $B$, $C$, $D$)?

We will consider the size of the whole figure as small, so that geodesic line segments within the figure are almost straight. We choose a point $E_1$ on geodesic $AD$, and a point $E_2$ on geodesic $BC$, both points near $E_0$. Then, the gap between geodesics $AD$ and $BC$ is determined by vector $\overrightarrow{E_1 E_2}$. To exclude ambiguity of the choice of points $E_1$ and $E_2$, this vector should be projected onto the plane perpendicular to both geodesics $AD$ and $BC$. 

We therefore choose two linearly independent unit vectors $\vec b_1$ and $\vec b_2$, both of them perpendicular to geodesics $AD$ and $BC$, and consider quantities
\begin{equation} \label{eq:app:xi12-def}
	\xi_1 = \vec b_1 \cdot \overrightarrow{E_1 E_2}, \quad
	\xi_2 = \vec b_2 \cdot \overrightarrow{E_1 E_2}.
\end{equation}
They completely quantify the distance between the geodesics $AD$ and $BC$.

It is convenient to calculate $\xi_1$ and $\xi_2$ in the Riemann normal coordinates~\cite[\S11.6]{Wheeler_book} with the origin at point $O$. 
The metric tensor $g_{ik}(\r)$ in these coordinates has the form of Eq.~(\ref{eq:2:g-series}), and geodesics $OAB$ and $OCD$ (that pass through the origin) are simply straight lines. If points $P$ and $Q$ lie within a small $\epsilon$-neighborhood of point $O$, then the segment of the geodesic line that joins two points $P$ and $Q$ obeys the following parametric equation in the normal coordinates~\cite{Brewin2009}:
\begin{multline} \label{eq:app:geodesics-1}
	r^i(\lambda) = (1-\lambda) r_P^i + \lambda r_Q^i \\
	+ \frac{\lambda(1-\lambda)}{3} R{^i}_{klm} (r_Q^k-r_P^k) r_P^l (r_Q^m-r_P^m) + {\cal O}(\epsilon^4),
\end{multline}
where parameter $\lambda$ is equal to 0 at $P$ and 1 at $Q$. 

Before applying Eq.~(\ref{eq:app:geodesics-1}), we slightly modify it with the aid of skew symmetry ($R{^i}_{klm} = -R{^i}_{kml}$) of the Riemann tensor. First, we notice that $R{^i}_{klm}r_P^lr_P^m = 0$ due to this skew symmetry, and hence the term $(-r_P^m)$ can be omitted in Eq.~(\ref{eq:app:geodesics-1}):
\begin{multline} \label{eq:app:geodesics-2}
	r^i(\lambda) = (1-\lambda) r_P^i + \lambda r_Q^i \\
	+ \frac{\lambda(1-\lambda)}{3} (R{^i}_{klm} r_Q^k r_P^l r_Q^m - R{^i}_{klm} r_P^k r_P^l r_Q^m) + {\cal O}(\epsilon^4).
\end{multline}
The term $R{^i}_{klm} r_Q^k r_P^l r_Q^m$ can be further transformed to $(-R{^i}_{klm} r_Q^k r_Q^l r_P^m)$ by swapping indices $l$ and $m$ and employing the skew symmetry:
\begin{multline} \label{eq:app:geodesics-3}
	r^i(\lambda) = (1-\lambda) r_P^i + \lambda r_Q^i \\
	- \frac{\lambda(1-\lambda)}{3} R{^i}_{klm} (r_Q^k r_Q^l r_P^m + r_P^k r_P^l r_Q^m) + {\cal O}(\epsilon^4).
\end{multline}
Later, we assume that the whole figure $OABCD$ belongs to the $\epsilon$-neighborhood of point $O$ with a small size $\epsilon$, and neglect the residue term ${\cal O}(\epsilon^4)$.

We describe figure $OABCD$ by two vectors $\vec a_1$, $\vec a_2$ (see Fig.~\ref{fig:quadrilateral_calc}) and two numbers $\mu_1$, $\mu_2$:
\begin{equation} \label{eq:app:a12-mu12-def}
	\vec a_1 = \overrightarrow{OA}, \quad
	\vec a_2 = \overrightarrow{OC}, \quad
	\mu_1 = \frac{|\overrightarrow{OB}|}{|\overrightarrow{OA}|}, \quad
	\mu_2 = \frac{|\overrightarrow{OD}|}{|\overrightarrow{OC}|},
\end{equation}
so that radius-vectors of points $A$, $B$, $C$, $D$ are
\begin{equation} \label{eq:app:r-ABCD}
	\r_A = \vec a_1, \quad
	\r_B = \mu_1 \vec a_1, \quad
	\r_C = \vec a_2, \quad
	\r_D = \mu_2 \vec a_2.
\end{equation}
Point $E_0$ lies at the intersection of \emph{straight lines} $AD$ and $BC$. Hence, its radius-vector expresses as
\begin{equation} \label{eq:app:r-E0}
	\r_{E_0} = (1-\lambda_1) \r_A + \lambda_1 \r_D = (1-\lambda_2) \r_C + \lambda_2 \r_B ,
\end{equation}
where $\lambda_1$ and $\lambda_2$ are some numbers. Substituting here $\r_A$, $\r_B$, $\r_C$ and $\r_D$ from Eq.~(\ref{eq:app:r-ABCD}), and collecting the terms with $\vec a_1$ and $\vec a_2$ separately, one can obtain two equations for $\lambda_1$ and $\lambda_2$:
\begin{equation} \label{eq:app:lambda12-equation}
	1-\lambda_1=\lambda_2\mu_1, \qquad
	\lambda_1\mu_2=1-\lambda_2.
\end{equation}
Solution of these equations is
\begin{equation} \label{eq:app:lambda12}
	\lambda_1 = \frac{\mu_1-1}{\mu_1\mu_2-1} , \qquad
	\lambda_2 = \frac{\mu_2-1}{\mu_1\mu_2-1} .
\end{equation}

Coordinates of point $E_1$ can be obtained by substitution of $\r_A = \vec a_1$, $\r_D = \mu_2 \vec a_2$ and $\lambda_1$ as $\r_P$, $\r_Q$ and $\lambda$ into Eq.~(\ref{eq:app:geodesics-3}). It is evident from comparison with Eq.~(\ref{eq:app:r-E0}) that the first term in the resulting expression is equal to $\r_{E_0}$. Hence,
\begin{equation} \label{eq:app:r-E1}
	\r_{E_1} = \r_{E_0} - \frac{\lambda_1(1-\lambda_1)}{3} \mu_2 R{^i}_{klm} (\mu_2 a_2^k a_2^l a_1^m + a_1^k a_1^l a_2^m) .
\end{equation}
Similarly, coordinates of point $E_2$ can be obtained by substitution of $\r_C = \vec a_2$, $\r_B = \mu_1 \vec a_1$ and $\lambda_2$ as $\r_P$, $\r_Q$ and $\lambda$ into Eq.~(\ref{eq:app:geodesics-3}):
\begin{equation} \label{eq:app:r-E2}
	\r_{E_2} = \r_{E_0} - \frac{\lambda_2(1-\lambda_2)}{3} \mu_1 R{^i}_{klm} (\mu_1 a_1^k a_1^l a_2^m + a_2^k a_2^l a_1^m) .
\end{equation}
The prefactors before $R{^i}_{klm}$ in equations~(\ref{eq:app:r-E1}) and~(\ref{eq:app:r-E2}) are equal to each other, that can be easily seen by expressing $(1-\lambda_1)$ and $(1-\lambda_2)$ via Eq.~(\ref{eq:app:lambda12-equation}):
\begin{equation} \label{eq:app:prefactors-are equal}
	\frac{\lambda_1(1-\lambda_1)}{3} \mu_2 = \frac{\lambda_2(1-\lambda_2)}{3} \mu_1 = \frac{\lambda_1\lambda_2\mu_1\mu_2}{3} \, .
\end{equation}
This fact simplifies calculation of vector $\overrightarrow{E_1 E_2} \equiv \r_{E_2} - \r_{E_1}$ and quantities
\begin{equation} \label{eq:app:xi12-again}
	\xi_1 = \vec b_1 \cdot (\r_{E_2} - \r_{E_1}) \;\; \text{and} \;\;
	\xi_2 = \vec b_2 \cdot (\r_{E_2} - \r_{E_1}).
\end{equation}

Collecting Eqs.~(\ref{eq:app:r-E1}) -- (\ref{eq:app:xi12-again}) together, and taking values of $\lambda_1$ and $\lambda_2$ from Eq.~(\ref{eq:app:lambda12}), one can find $\xi_1$ and~$\xi_2$ as
\begin{multline} \label{eq:app:xi1-result}
	\xi_1 = \frac{\mu_1\mu_2(1-\mu_1)(1-\mu_2)}{3(\mu_1\mu_2-1)^2} R_{iklm} \, b_1^i \\
	\times \left[ (\mu_2-1) a_2^k a_2^l a_1^m - (\mu_1-1) a_1^k a_1^l a_2^m \right]  + {\cal O}(\epsilon^4)
\end{multline}
and
\begin{multline} \label{eq:app:xi2-result}
	\xi_2 = \frac{\mu_1\mu_2(1-\mu_1)(1-\mu_2)}{3(\mu_1\mu_2-1)^2} R_{iklm} \, b_2^i \\
	\times \left[ (\mu_2-1) a_2^k a_2^l a_1^m - (\mu_1-1) a_1^k a_1^l a_2^m \right]  + {\cal O}(\epsilon^4).
\end{multline}

Equations~(\ref{eq:app:xi1-result}) and~(\ref{eq:app:xi2-result}) solve the problem of calculating the gap between geodesics $AD$ and $BC$ (described by $\xi_1$ and $\xi_2$) expressed through parameters $\vec a_1$, $\vec a_2$, $\mu_1$, $\mu_2$ of figure $OABCD$, unit vectors $\vec b_1$, $\vec b_2$ perpendicular to the figure, and Riemann curvature tensor $R_{iklm}$ at point $O$. Though we have obtained these equations with a special choice of coordinates, they are written in an invariant form, and therefore are valid in any coordinate frame.

It is evident from Eqs.~(\ref{eq:app:xi1-result}) and~(\ref{eq:app:xi2-result}) that the quantities $\xi_1$ and $\xi_2$, that characterize the distance between geodesics $AD$ and $BC$, are proportional to some linear combinations of components of curvature tensor $R_{iklm}$. Measuring the gaps between geodesics $AD$ and $BC$ in different figures $OABCD$, one can thus restore some information about tensor $R_{iklm}$. Let us now figure out, how much information about $R_{iklm}$ can be obtained in this way. 

For this goal, it is enough to fix the values of $\mu_1$ and $\mu_2$ at
\begin{equation}
	\mu_1 = \mu_2 = 2,
\end{equation}
i.~e. to restrict our attention to those figures, in which $|\overrightarrow{OA}| = |\overrightarrow{AB}|$ and $|\overrightarrow{OC}| = |\overrightarrow{CD}|$. In this case, Eqs.~(\ref{eq:app:xi1-result}) and~(\ref{eq:app:xi2-result}) are simplified to
\begin{equation} \label{eq:app:xi1-simplified}
	\xi_1 = \frac{4}{27} R_{iklm} \, b_1^i \left( a_2^k a_2^l a_1^m - a_1^k a_1^l a_2^m \right)
\end{equation}
and
\begin{equation} \label{eq:app:xi2-simplified}
	\xi_2 = \frac{4}{27} R_{iklm} \, b_2^i \left( a_2^k a_2^l a_1^m - a_1^k a_1^l a_2^m \right)
\end{equation}
(we have also omitted the residue term ${\cal O}(\epsilon^4)$). In particular, choosing a small length scale $\epsilon$ and setting
\begin{equation}
	\vec a_1 = \epsilon \vec e_x, \quad
	\vec a_2 = \epsilon \vec e_y, \quad
	\vec b_1 = \vec e_z, \quad
	\vec b_2 = \vec e_t,
\end{equation}
where $\vec e_x$, $\vec e_y$, $\vec e_z$ and $\vec e_t$ are unit vectors along the coordinate axes, one obtains
\begin{equation}
	\xi_1 = \frac{4\epsilon^3}{27} \left( R_{zyyx} - R_{zxxy} \right), \quad
	\xi_2 = \frac{4\epsilon^3}{27} \left( R_{tyyx} - R_{txxy} \right). \quad
\end{equation}
Similarly, setting
\begin{equation}
	\vec a_1 = \epsilon \vec e_x, \quad
	\vec a_2 = -\epsilon \vec e_y, \quad
	\vec b_1 = \vec e_z, \quad
	\vec b_2 = \vec e_t,
\end{equation}
one can get
\begin{equation}
	\xi_1 = \frac{4\epsilon^3}{27} \left( R_{zyyx} + R_{zxxy} \right), \quad
	\xi_2 = \frac{4\epsilon^3}{27} \left( R_{tyyx} + R_{txxy} \right). \quad
\end{equation}
Hence, from measurement of quantities $\xi_1$ and $\xi_2$ for the two different quadrilaterals, one can restore four components of the curvature tensor:
\begin{equation*}
	R_{zyyx}, \quad
	R_{zxxy}, \quad
	R_{tyyx}, \quad \text{and} \;\;\;
	R_{txxy}.
\end{equation*}
In the same manner, measuring other quadrilaterals, with geodesics $OAB$ and $OCD$ directing along other coordinate axes, one can obtain all curvature tensor components of form $R_{ikkl}$, where $i \neq k$, $i \neq l$ and $k \neq l$. Some of these components are equal to each other due to symmetries of the Riemann tensor, namely $R_{ikkl} = R_{lkki}$. There are 12 algebraically independent components of this form:
\begin{eqnarray*}
	R_{xtty}, \quad
	R_{xttz}, \quad
	R_{yttz}, \quad
	R_{txxy}, \quad
	R_{txxz}, \quad
	R_{yxxz}, \\
	R_{tyyx}, \quad
	R_{tyyz}, \quad
	R_{xyyz}, \quad
	R_{tzzx}, \quad
	R_{tzzy}, \quad
	R_{xzzy}.
\end{eqnarray*}

Hence, \emph{at least 12 independent components of curvature tensor $R_{iklm}$ can be obtained from measuring the gaps between geodesics $OAB$ and $OCD$ in different quadrilaterals $OABCD$.}

But what about other 8 components? One can study more orientations of quadrilaterals on the subject of extracting more information about the curvature tensor. Instead, we will address much simpler considerations based on group theory.

All possible values of the Riemann curvature tensor $R_{iklm}$ consist a 20-dimensional representation of the Lorentz group. It can be decomposed into three \emph{irreducible} representations: a 10-dimensional one that describes possible values of the Weyl tensor, a 9-dimensional one related to the traceless part of the Ricci tensor, and a one-dimensional representation for the scalar curvature~\cite{Strichartz1988}. Correspondingly, all 20 degrees of freedom of the Riemann tensor consist of $10+9+1$ degrees of freedom related to the Weyl tensor, the traceless part of the Ricci tensor, and the scalar curvature. Each of these three pieces either can be completely restored from measurements on different quadrilaterals, or cannot be restored at all in such a way.

Suppose that the Weyl tensor cannot be restored from measuring the gaps between geodesics in different quadrilaterals. Then, no more than 10 degrees of freedom of the Riemann tensor $R_{iklm}$ can be restored---namely, all 20 degrees of freedom of $R_{iklm}$ minus 10 degrees of freedom of the Weyl tensor \emph{at most}. But this contradicts to the above-mentioned statement that \emph{at least} 12 independent components of $R_{iklm}$ can be obtained in such a way. Hence, the argument by contradiction leads us to the conclusion that \emph{the Weyl tensor can be restored from measuring the gaps between geodesics in different quadrilaterals $OABCD$.}

The same argument can be applied to the traceless part of the Ricci tensor, that leads to the following conclusion: \emph{the traceless part of the Ricci tensor also can be restored from measuring the gaps between geodesics in quadrilaterals.}

The only remaining question is whether the scalar curvature can be restored in a similar way. If we suppose that it can, then the whole Riemann tensor $R_{iklm}$ could be obtained from measuring the gaps in quadrilaterals. Therefore we could distinguish between the flat spacetime, where $R_{iklm}=0$ and a spacetime of a constant nonzero sectional curvature (de Sitter or anti-de Sitter), where $R_{iklm}\neq0$, just by measuring the gaps in quadrilaterals. But it is impossible, since there is \emph{no gap} between geodesics $AD$ and $BC$, as in the flat spacetime, as in the de Sitter/anti-de Sitter one (see Section~\ref{sec:discussion}). Again, we come to a contradiction, and thus prove that \emph{the scalar curvature cannot be restored from measuring the gaps between geodesics in quadrilaterals.}

To summarize, the gap between geodesics $AD$ and $BC$ in the configuration $OABCD$ shown in Fig.~\ref{fig:quadrilateral_calc} (a complete quadrilateral) is characterized by two quantities $\xi_1 = \vec b_1 \cdot \overrightarrow{E_1 E_2}$ and $\xi_2 = \vec b_2 \cdot \overrightarrow{E_1 E_2}$. Equations~(\ref{eq:app:xi1-result}) and~(\ref{eq:app:xi2-result}) express these quantities via parameters of the quadrilateral and the local Riemann tensor $R_{iklm}$. By measuring $\xi_1$ and $\xi_2$ for different quadrilaterals around point $O$, one can get 19 of 20 parameters of the Riemann tensor at $O$: namely, 10 independent components of the Weyl tensor and 9 independent components of the traceless part of the Ricci tensor.


\bibliography{connection}

\begin{thebibliography}{58}%
\makeatletter
\providecommand \@ifxundefined [1]{%
 \@ifx{#1\undefined}
}%
\providecommand \@ifnum [1]{%
 \ifnum #1\expandafter \@firstoftwo
 \else \expandafter \@secondoftwo
 \fi
}%
\providecommand \@ifx [1]{%
 \ifx #1\expandafter \@firstoftwo
 \else \expandafter \@secondoftwo
 \fi
}%
\providecommand \natexlab [1]{#1}%
\providecommand \enquote  [1]{``#1''}%
\providecommand \bibnamefont  [1]{#1}%
\providecommand \bibfnamefont [1]{#1}%
\providecommand \citenamefont [1]{#1}%
\providecommand \href@noop [0]{\@secondoftwo}%
\providecommand \href [0]{\begingroup \@sanitize@url \@href}%
\providecommand \@href[1]{\@@startlink{#1}\@@href}%
\providecommand \@@href[1]{\endgroup#1\@@endlink}%
\providecommand \@sanitize@url [0]{\catcode `\\12\catcode `\$12\catcode
  `\&12\catcode `\#12\catcode `\^12\catcode `\_12\catcode `\%12\relax}%
\providecommand \@@startlink[1]{}%
\providecommand \@@endlink[0]{}%
\providecommand \url  [0]{\begingroup\@sanitize@url \@url }%
\providecommand \@url [1]{\endgroup\@href {#1}{\urlprefix }}%
\providecommand \urlprefix  [0]{URL }%
\providecommand \Eprint [0]{\href }%
\providecommand \doibase [0]{https://doi.org/}%
\providecommand \selectlanguage [0]{\@gobble}%
\providecommand \bibinfo  [0]{\@secondoftwo}%
\providecommand \bibfield  [0]{\@secondoftwo}%
\providecommand \translation [1]{[#1]}%
\providecommand \BibitemOpen [0]{}%
\providecommand \bibitemStop [0]{}%
\providecommand \bibitemNoStop [0]{.\EOS\space}%
\providecommand \EOS [0]{\spacefactor3000\relax}%
\providecommand \BibitemShut  [1]{\csname bibitem#1\endcsname}%
\let\auto@bib@innerbib\@empty
\bibitem [{\citenamefont {Misner}\ \emph {et~al.}(1973)\citenamefont {Misner},
  \citenamefont {Thorne},\ and\ \citenamefont {Wheeler}}]{Wheeler_book}%
  \BibitemOpen
  \bibfield  {author} {\bibinfo {author} {\bibfnamefont {C.~W.}\ \bibnamefont
  {Misner}}, \bibinfo {author} {\bibfnamefont {K.~S.}\ \bibnamefont {Thorne}},\
  and\ \bibinfo {author} {\bibfnamefont {J.~A.}\ \bibnamefont {Wheeler}},\
  }\href@noop {} {\emph {\bibinfo {title} {Gravitation}}}\ (\bibinfo
  {publisher} {W. H. Freeman and Company},\ \bibinfo {year} {1973})\BibitemShut
  {NoStop}%
\bibitem [{\citenamefont {Synge}(1960)}]{Synge_book}%
  \BibitemOpen
  \bibfield  {author} {\bibinfo {author} {\bibfnamefont {J.}~\bibnamefont
  {Synge}},\ }\href {https://books.google.de/books?id=06kNAQAAIAAJ} {\emph
  {\bibinfo {title} {Relativity: The General Theory}}}\ (\bibinfo  {publisher}
  {North-Holland Publishing Company},\ \bibinfo {year} {1960})\BibitemShut
  {NoStop}%
\bibitem [{\citenamefont {{J. Aasi {\it et al.} (LIGO Scientific
  Collaboration)}}(2015)}]{Aasi2015LIGO}%
  \BibitemOpen
  \bibfield  {author} {\bibinfo {author} {\bibnamefont {{J. Aasi {\it et al.}
  (LIGO Scientific Collaboration)}}},\ }\bibfield  {title} {\bibinfo {title}
  {{Advanced LIGO}},\ }\href {https://doi.org/10.1088/0264-9381/32/7/074001}
  {\bibfield  {journal} {\bibinfo  {journal} {Classical and Quantum Gravity}\
  }\textbf {\bibinfo {volume} {32}},\ \bibinfo {pages} {074001} (\bibinfo
  {year} {2015})}\BibitemShut {NoStop}%
\bibitem [{\citenamefont {{F. Acernese {\it et al.} (Virgo
  Collaboration)}}(2015)}]{Acernese2015Virgo}%
  \BibitemOpen
  \bibfield  {author} {\bibinfo {author} {\bibnamefont {{F. Acernese {\it et
  al.} (Virgo Collaboration)}}},\ }\bibfield  {title} {\bibinfo {title}
  {{Advanced Virgo: a second-generation interferometric gravitational wave
  detector}},\ }\href {https://doi.org/10.1088/0264-9381/32/2/024001}
  {\bibfield  {journal} {\bibinfo  {journal} {Classical and Quantum Gravity}\
  }\textbf {\bibinfo {volume} {32}},\ \bibinfo {pages} {024001} (\bibinfo
  {year} {2015})}\BibitemShut {NoStop}%
\bibitem [{\citenamefont {{T. Akutsu {\it et al.} (KAGRA
  Collaboration)}}(2019)}]{Akutsu2019KAGRA}%
  \BibitemOpen
  \bibfield  {author} {\bibinfo {author} {\bibnamefont {{T. Akutsu {\it et al.}
  (KAGRA Collaboration)}}},\ }\bibfield  {title} {\bibinfo {title} {{KAGRA: 2.5
  generation interferometric gravitational wave detector}},\ }\href
  {https://doi.org/10.1038/s41550-018-0658-y} {\bibfield  {journal} {\bibinfo
  {journal} {Nature Astronomy}\ }\textbf {\bibinfo {volume} {3}},\ \bibinfo
  {pages} {35} (\bibinfo {year} {2019})}\BibitemShut {NoStop}%
\bibitem [{\citenamefont {{P. Amaro-Seoane {\it et al.} (LISA Consortium
  Collaboration)}}(2017)}]{Amaro-Seoane2017LISA}%
  \BibitemOpen
  \bibfield  {author} {\bibinfo {author} {\bibnamefont {{P. Amaro-Seoane {\it
  et al.} (LISA Consortium Collaboration)}}},\ }\href
  {https://doi.org/10.48550/ARXIV.1702.00786} {\bibinfo {title} {{Laser
  Interferometer Space Antenna, arXiv:1702.00786}}} (\bibinfo {year}
  {2017})\BibitemShut {NoStop}%
\bibitem [{\citenamefont {Blas}\ and\ \citenamefont
  {Jenkins}(2022)}]{Blas2022PRL}%
  \BibitemOpen
  \bibfield  {author} {\bibinfo {author} {\bibfnamefont {D.}~\bibnamefont
  {Blas}}\ and\ \bibinfo {author} {\bibfnamefont {A.~C.}\ \bibnamefont
  {Jenkins}},\ }\bibfield  {title} {\bibinfo {title} {Bridging the
  $\ensuremath{\mu}\mathrm{Hz}$ gap in the gravitational-wave landscape with
  binary resonances},\ }\href {https://doi.org/10.1103/PhysRevLett.128.101103}
  {\bibfield  {journal} {\bibinfo  {journal} {Phys. Rev. Lett.}\ }\textbf
  {\bibinfo {volume} {128}},\ \bibinfo {pages} {101103} (\bibinfo {year}
  {2022})}\BibitemShut {NoStop}%
\bibitem [{\citenamefont {Hohensee}\ \emph {et~al.}(2012)\citenamefont
  {Hohensee}, \citenamefont {Estey}, \citenamefont {Hamilton}, \citenamefont
  {Zeilinger},\ and\ \citenamefont {M\"uller}}]{Hohensee2012}%
  \BibitemOpen
  \bibfield  {author} {\bibinfo {author} {\bibfnamefont {M.~A.}\ \bibnamefont
  {Hohensee}}, \bibinfo {author} {\bibfnamefont {B.}~\bibnamefont {Estey}},
  \bibinfo {author} {\bibfnamefont {P.}~\bibnamefont {Hamilton}}, \bibinfo
  {author} {\bibfnamefont {A.}~\bibnamefont {Zeilinger}},\ and\ \bibinfo
  {author} {\bibfnamefont {H.}~\bibnamefont {M\"uller}},\ }\bibfield  {title}
  {\bibinfo {title} {Force-free gravitational redshift: Proposed gravitational
  {Aharonov-Bohm} experiment},\ }\href
  {https://doi.org/10.1103/PhysRevLett.108.230404} {\bibfield  {journal}
  {\bibinfo  {journal} {Phys. Rev. Lett.}\ }\textbf {\bibinfo {volume} {108}},\
  \bibinfo {pages} {230404} (\bibinfo {year} {2012})}\BibitemShut {NoStop}%
\bibitem [{\citenamefont {Overstreet}\ \emph {et~al.}(2022)\citenamefont
  {Overstreet}, \citenamefont {Asenbaum}, \citenamefont {Curti}, \citenamefont
  {Kim},\ and\ \citenamefont {Kasevich}}]{Overstreet2022}%
  \BibitemOpen
  \bibfield  {author} {\bibinfo {author} {\bibfnamefont {C.}~\bibnamefont
  {Overstreet}}, \bibinfo {author} {\bibfnamefont {P.}~\bibnamefont
  {Asenbaum}}, \bibinfo {author} {\bibfnamefont {J.}~\bibnamefont {Curti}},
  \bibinfo {author} {\bibfnamefont {M.}~\bibnamefont {Kim}},\ and\ \bibinfo
  {author} {\bibfnamefont {M.~A.}\ \bibnamefont {Kasevich}},\ }\bibfield
  {title} {\bibinfo {title} {Observation of a gravitational {Aharonov-Bohm}
  effect},\ }\href {https://doi.org/10.1126/science.abl7152} {\bibfield
  {journal} {\bibinfo  {journal} {Science}\ }\textbf {\bibinfo {volume}
  {375}},\ \bibinfo {pages} {226} (\bibinfo {year} {2022})}\BibitemShut
  {NoStop}%
\bibitem [{\citenamefont {Roura}(2022)}]{Roura2022}%
  \BibitemOpen
  \bibfield  {author} {\bibinfo {author} {\bibfnamefont {A.}~\bibnamefont
  {Roura}},\ }\bibfield  {title} {\bibinfo {title} {Quantum probe of space-time
  curvature},\ }\href {https://doi.org/10.1126/science.abm6854} {\bibfield
  {journal} {\bibinfo  {journal} {Science}\ }\textbf {\bibinfo {volume}
  {375}},\ \bibinfo {pages} {142} (\bibinfo {year} {2022})}\BibitemShut
  {NoStop}%
\bibitem [{\citenamefont {D'Inverno}(1992)}]{dInverno_book}%
  \BibitemOpen
  \bibfield  {author} {\bibinfo {author} {\bibfnamefont {R.}~\bibnamefont
  {D'Inverno}},\ }\href {https://books.google.de/books?id=yqcT3VICuvMC} {\emph
  {\bibinfo {title} {Introducing Einstein's Relativity}}}\ (\bibinfo
  {publisher} {Clarendon Press},\ \bibinfo {year} {1992})\BibitemShut {NoStop}%
\bibitem [{\citenamefont {Landau}\ and\ \citenamefont {Lifshitz}(2000)}]{LL2}%
  \BibitemOpen
  \bibfield  {author} {\bibinfo {author} {\bibfnamefont {L.~D.}\ \bibnamefont
  {Landau}}\ and\ \bibinfo {author} {\bibfnamefont {E.~M.}\ \bibnamefont
  {Lifshitz}},\ }\href {https://books.google.de/books?id=X18PF4oKyrUC} {\emph
  {\bibinfo {title} {The Classical Theory of Fields: Volume 2}}},\ Course of
  theoretical physics\ (\bibinfo  {publisher} {Butterworth-Heinemann},\
  \bibinfo {year} {2000})\BibitemShut {NoStop}%
\bibitem [{\citenamefont {Dittus}\ \emph {et~al.}(2008)\citenamefont {Dittus},
  \citenamefont {L{\"a}mmerzahl},\ and\ \citenamefont
  {Turyshev}}]{Lasers_Clocks_book}%
  \BibitemOpen
  \bibinfo {editor} {\bibfnamefont {H.}~\bibnamefont {Dittus}}, \bibinfo
  {editor} {\bibfnamefont {C.}~\bibnamefont {L{\"a}mmerzahl}},\ and\ \bibinfo
  {editor} {\bibfnamefont {S.}~\bibnamefont {Turyshev}},\ eds.,\ \href
  {https://books.google.de/books?id=pGRROFE5zRoC} {\emph {\bibinfo {title}
  {Lasers, Clocks and Drag-Free Control: Exploration of Relativistic Gravity in
  Space}}},\ Astrophysics and Space Science Library\ (\bibinfo  {publisher}
  {Springer Berlin Heidelberg},\ \bibinfo {year} {2008})\BibitemShut {NoStop}%
\bibitem [{\citenamefont {Hilbert}\ and\ \citenamefont
  {Cohn-Vossen}(1999)}]{Hilbert_book}%
  \BibitemOpen
  \bibfield  {author} {\bibinfo {author} {\bibfnamefont {D.}~\bibnamefont
  {Hilbert}}\ and\ \bibinfo {author} {\bibfnamefont {S.}~\bibnamefont
  {Cohn-Vossen}},\ }\href {https://books.google.de/books?id=7WY5AAAAQBAJ}
  {\emph {\bibinfo {title} {Geometry and the Imagination}}},\ AMS Chelsea
  Publishing Series\ (\bibinfo  {publisher} {AMS Chelsea Pub.},\ \bibinfo
  {year} {1999})\BibitemShut {NoStop}%
\bibitem [{\citenamefont {Dribus}(2017)}]{Dribus_book}%
  \BibitemOpen
  \bibfield  {author} {\bibinfo {author} {\bibfnamefont {B.}~\bibnamefont
  {Dribus}},\ }\href {https://books.google.de/books?id=u\_q\_DgAAQBAJ} {\emph
  {\bibinfo {title} {Discrete Causal Theory: Emergent Spacetime and the Causal
  Metric Hypothesis}}}\ (\bibinfo  {publisher} {Springer International
  Publishing},\ \bibinfo {year} {2017})\BibitemShut {NoStop}%
\bibitem [{\citenamefont {Surya}(2019)}]{Surya2019}%
  \BibitemOpen
  \bibfield  {author} {\bibinfo {author} {\bibfnamefont {S.}~\bibnamefont
  {Surya}},\ }\bibfield  {title} {\bibinfo {title} {The causal set approach to
  quantum gravity},\ }\href {https://doi.org/10.1007/s41114-019-0023-1}
  {\bibfield  {journal} {\bibinfo  {journal} {Living Reviews in Relativity}\
  }\textbf {\bibinfo {volume} {22}},\ \bibinfo {pages} {5} (\bibinfo {year}
  {2019})}\BibitemShut {NoStop}%
\bibitem [{\citenamefont {Loll}(2019)}]{Loll2020}%
  \BibitemOpen
  \bibfield  {author} {\bibinfo {author} {\bibfnamefont {R.}~\bibnamefont
  {Loll}},\ }\bibfield  {title} {\bibinfo {title} {Quantum gravity from causal
  dynamical triangulations: a review},\ }\href
  {https://doi.org/10.1088/1361-6382/ab57c7} {\bibfield  {journal} {\bibinfo
  {journal} {Classical and Quantum Gravity}\ }\textbf {\bibinfo {volume}
  {37}},\ \bibinfo {pages} {013002} (\bibinfo {year} {2019})}\BibitemShut
  {NoStop}%
\bibitem [{\citenamefont {Benincasa}\ and\ \citenamefont
  {Dowker}(2010)}]{Benincasa2010}%
  \BibitemOpen
  \bibfield  {author} {\bibinfo {author} {\bibfnamefont {D.~M.~T.}\
  \bibnamefont {Benincasa}}\ and\ \bibinfo {author} {\bibfnamefont
  {F.}~\bibnamefont {Dowker}},\ }\bibfield  {title} {\bibinfo {title} {Scalar
  curvature of a causal set},\ }\href
  {https://doi.org/10.1103/PhysRevLett.104.181301} {\bibfield  {journal}
  {\bibinfo  {journal} {Phys. Rev. Lett.}\ }\textbf {\bibinfo {volume} {104}},\
  \bibinfo {pages} {181301} (\bibinfo {year} {2010})}\BibitemShut {NoStop}%
\bibitem [{\citenamefont {Hawking}\ \emph {et~al.}(1976)\citenamefont
  {Hawking}, \citenamefont {King},\ and\ \citenamefont
  {McCarthy}}]{Hawking1976}%
  \BibitemOpen
  \bibfield  {author} {\bibinfo {author} {\bibfnamefont {S.~W.}\ \bibnamefont
  {Hawking}}, \bibinfo {author} {\bibfnamefont {A.~R.}\ \bibnamefont {King}},\
  and\ \bibinfo {author} {\bibfnamefont {P.~J.}\ \bibnamefont {McCarthy}},\
  }\bibfield  {title} {\bibinfo {title} {A new topology for curved space–time
  which incorporates the causal, differential, and conformal structures},\
  }\href {https://doi.org/10.1063/1.522874} {\bibfield  {journal} {\bibinfo
  {journal} {Journal of Mathematical Physics}\ }\textbf {\bibinfo {volume}
  {17}},\ \bibinfo {pages} {174} (\bibinfo {year} {1976})}\BibitemShut
  {NoStop}%
\bibitem [{\citenamefont {Malament}(1977)}]{Malament1977}%
  \BibitemOpen
  \bibfield  {author} {\bibinfo {author} {\bibfnamefont {D.~B.}\ \bibnamefont
  {Malament}},\ }\bibfield  {title} {\bibinfo {title} {The class of continuous
  timelike curves determines the topology of spacetime},\ }\href
  {https://doi.org/10.1063/1.523436} {\bibfield  {journal} {\bibinfo  {journal}
  {Journal of Mathematical Physics}\ }\textbf {\bibinfo {volume} {18}},\
  \bibinfo {pages} {1399} (\bibinfo {year} {1977})}\BibitemShut {NoStop}%
\bibitem [{\citenamefont {Topalov}\ and\ \citenamefont
  {Matveev}(2003)}]{Topalov-Matveev2003}%
  \BibitemOpen
  \bibfield  {author} {\bibinfo {author} {\bibfnamefont {P.}~\bibnamefont
  {Topalov}}\ and\ \bibinfo {author} {\bibfnamefont {V.~S.}\ \bibnamefont
  {Matveev}},\ }\bibfield  {title} {\bibinfo {title} {Geodesic equivalence via
  integrability},\ }\href {https://doi.org/10.1023/A:1022166218282} {\bibfield
  {journal} {\bibinfo  {journal} {Geometriae Dedicata}\ }\textbf {\bibinfo
  {volume} {96}},\ \bibinfo {pages} {91} (\bibinfo {year} {2003})}\BibitemShut
  {NoStop}%
\bibitem [{\citenamefont {Matveev}(2012)}]{Matveev2012}%
  \BibitemOpen
  \bibfield  {author} {\bibinfo {author} {\bibfnamefont {V.~S.}\ \bibnamefont
  {Matveev}},\ }\bibfield  {title} {\bibinfo {title} {Geodesically equivalent
  metrics in general relativity},\ }\href
  {https://doi.org/https://doi.org/10.1016/j.geomphys.2011.04.019} {\bibfield
  {journal} {\bibinfo  {journal} {Journal of Geometry and Physics}\ }\textbf
  {\bibinfo {volume} {62}},\ \bibinfo {pages} {675} (\bibinfo {year}
  {2012})}\BibitemShut {NoStop}%
\bibitem [{\citenamefont {Ehlers}\ \emph {et~al.}(2012)\citenamefont {Ehlers},
  \citenamefont {Pirani},\ and\ \citenamefont {Schild}}]{Ehlers2012}%
  \BibitemOpen
  \bibfield  {author} {\bibinfo {author} {\bibfnamefont {J.}~\bibnamefont
  {Ehlers}}, \bibinfo {author} {\bibfnamefont {F.~A.~E.}\ \bibnamefont
  {Pirani}},\ and\ \bibinfo {author} {\bibfnamefont {A.}~\bibnamefont
  {Schild}},\ }\bibfield  {title} {\bibinfo {title} {{Republication of: The
  geometry of free fall and light propagation}},\ }\href
  {https://doi.org/10.1007/s10714-012-1353-4} {\bibfield  {journal} {\bibinfo
  {journal} {General Relativity and Gravitation}\ }\textbf {\bibinfo {volume}
  {44}},\ \bibinfo {pages} {1587} (\bibinfo {year} {2012})}\BibitemShut
  {NoStop}%
\bibitem [{\citenamefont {Matveev}\ and\ \citenamefont
  {Scholz}(2020)}]{Matveev-Scholz2020}%
  \BibitemOpen
  \bibfield  {author} {\bibinfo {author} {\bibfnamefont {V.~S.}\ \bibnamefont
  {Matveev}}\ and\ \bibinfo {author} {\bibfnamefont {E.}~\bibnamefont
  {Scholz}},\ }\bibfield  {title} {\bibinfo {title} {Light cone and {W}eyl
  compatibility of conformal and projective structures},\ }\href
  {https://doi.org/10.1007/s10714-020-02716-9} {\bibfield  {journal} {\bibinfo
  {journal} {General Relativity and Gravitation}\ }\textbf {\bibinfo {volume}
  {52}},\ \bibinfo {pages} {66} (\bibinfo {year} {2020})}\BibitemShut {NoStop}%
\bibitem [{\citenamefont
  {Desloge}(1989{\natexlab{a}})}]{Desloge1989metrosphere}%
  \BibitemOpen
  \bibfield  {author} {\bibinfo {author} {\bibfnamefont {E.~A.}\ \bibnamefont
  {Desloge}},\ }\bibfield  {title} {\bibinfo {title} {A theoretical device for
  space and time measurements},\ }\href {https://doi.org/10.1007/BF00731879}
  {\bibfield  {journal} {\bibinfo  {journal} {Foundations of Physics}\ }\textbf
  {\bibinfo {volume} {19}},\ \bibinfo {pages} {1191} (\bibinfo {year}
  {1989}{\natexlab{a}})}\BibitemShut {NoStop}%
\bibitem [{\citenamefont {Desloge}(1989{\natexlab{b}})}]{Desloge1989clock}%
  \BibitemOpen
  \bibfield  {author} {\bibinfo {author} {\bibfnamefont {E.~A.}\ \bibnamefont
  {Desloge}},\ }\bibfield  {title} {\bibinfo {title} {{A simple variation of
  the Marzke-Wheeler clock}},\ }\href {https://doi.org/10.1007/BF00759077}
  {\bibfield  {journal} {\bibinfo  {journal} {General Relativity and
  Gravitation}\ }\textbf {\bibinfo {volume} {21}},\ \bibinfo {pages} {677}
  (\bibinfo {year} {1989}{\natexlab{b}})}\BibitemShut {NoStop}%
\bibitem [{\citenamefont {Eisenhart}(1997)}]{Eisenhart_book}%
  \BibitemOpen
  \bibfield  {author} {\bibinfo {author} {\bibfnamefont {L.}~\bibnamefont
  {Eisenhart}},\ }\href {https://books.google.de/books?id=u9ukzVYtoNgC} {\emph
  {\bibinfo {title} {Riemannian Geometry}}},\ Princeton Landmarks in
  Mathematics and Physics\ (\bibinfo  {publisher} {Princeton University
  Press},\ \bibinfo {year} {1997})\BibitemShut {NoStop}%
\bibitem [{\citenamefont {Hawking}\ and\ \citenamefont
  {Ellis}(1973)}]{Hawking_book}%
  \BibitemOpen
  \bibfield  {author} {\bibinfo {author} {\bibfnamefont {S.}~\bibnamefont
  {Hawking}}\ and\ \bibinfo {author} {\bibfnamefont {G.}~\bibnamefont
  {Ellis}},\ }\href {https://books.google.de/books?id=QagG\_KI7Ll8C} {\emph
  {\bibinfo {title} {The Large Scale Structure of Space-Time}}},\ Cambridge
  Monographs on Mathematical Physics\ (\bibinfo  {publisher} {Cambridge
  University Press},\ \bibinfo {year} {1973})\BibitemShut {NoStop}%
\bibitem [{\citenamefont {Khriplovich}\ and\ \citenamefont
  {Rudenko}(2013)}]{Khriplovich2013}%
  \BibitemOpen
  \bibfield  {author} {\bibinfo {author} {\bibfnamefont {I.~B.}\ \bibnamefont
  {Khriplovich}}\ and\ \bibinfo {author} {\bibfnamefont {A.~S.}\ \bibnamefont
  {Rudenko}},\ }\bibfield  {title} {\bibinfo {title} {Gravitational
  four-fermion interaction and dynamics of the early {U}niverse},\ }\href
  {https://doi.org/10.1007/JHEP11(2013)174} {\bibfield  {journal} {\bibinfo
  {journal} {J. High Energy Phys.}\ }\textbf {\bibinfo {volume} {2013}}\bibinfo
   {number} { (11)},\ \bibinfo {pages} {174}}\BibitemShut {NoStop}%
\bibitem [{\citenamefont {Magueijo}\ \emph {et~al.}(2013)\citenamefont
  {Magueijo}, \citenamefont {Zlosnik},\ and\ \citenamefont
  {Kibble}}]{Magueijo2013}%
  \BibitemOpen
\bibfield  {number} {  }\bibfield  {author} {\bibinfo {author} {\bibfnamefont
  {J.}~\bibnamefont {Magueijo}}, \bibinfo {author} {\bibfnamefont {T.~G.}\
  \bibnamefont {Zlosnik}},\ and\ \bibinfo {author} {\bibfnamefont {T.~W.~B.}\
  \bibnamefont {Kibble}},\ }\bibfield  {title} {\bibinfo {title} {Cosmology
  with a spin},\ }\href {https://doi.org/10.1103/PhysRevD.87.063504} {\bibfield
   {journal} {\bibinfo  {journal} {Phys. Rev. D}\ }\textbf {\bibinfo {volume}
  {87}},\ \bibinfo {pages} {063504} (\bibinfo {year} {2013})}\BibitemShut
  {NoStop}%
\bibitem [{\citenamefont {Boos}\ and\ \citenamefont {Hehl}(2017)}]{Boos2017}%
  \BibitemOpen
  \bibfield  {author} {\bibinfo {author} {\bibfnamefont {J.}~\bibnamefont
  {Boos}}\ and\ \bibinfo {author} {\bibfnamefont {F.~W.}\ \bibnamefont
  {Hehl}},\ }\bibfield  {title} {\bibinfo {title} {Gravity-induced four-fermion
  contact interaction implies gravitational intermediate {W} and {Z} type gauge
  bosons},\ }\href {https://doi.org/10.1007/s10773-016-3216-3} {\bibfield
  {journal} {\bibinfo  {journal} {Int. J. Theor. Phys.}\ }\textbf {\bibinfo
  {volume} {56}},\ \bibinfo {pages} {751} (\bibinfo {year} {2017})}\BibitemShut
  {NoStop}%
\bibitem [{\citenamefont {Fradkin}\ and\ \citenamefont
  {Tseytlin}(1985)}]{Fradkin1985}%
  \BibitemOpen
  \bibfield  {author} {\bibinfo {author} {\bibfnamefont {E.}~\bibnamefont
  {Fradkin}}\ and\ \bibinfo {author} {\bibfnamefont {A.}~\bibnamefont
  {Tseytlin}},\ }\bibfield  {title} {\bibinfo {title} {Conformal
  supergravity},\ }\href
  {https://doi.org/https://doi.org/10.1016/0370-1573(85)90138-3} {\bibfield
  {journal} {\bibinfo  {journal} {Physics Reports}\ }\textbf {\bibinfo {volume}
  {119}},\ \bibinfo {pages} {233} (\bibinfo {year} {1985})}\BibitemShut
  {NoStop}%
\bibitem [{\citenamefont {Berkovits}\ and\ \citenamefont
  {Witten}(2004)}]{Berkovits2004}%
  \BibitemOpen
  \bibfield  {author} {\bibinfo {author} {\bibfnamefont {N.}~\bibnamefont
  {Berkovits}}\ and\ \bibinfo {author} {\bibfnamefont {E.}~\bibnamefont
  {Witten}},\ }\bibfield  {title} {\bibinfo {title} {Conformal supergravity in
  twistor-string theory},\ }\href
  {https://doi.org/10.1088/1126-6708/2004/08/009} {\bibfield  {journal}
  {\bibinfo  {journal} {Journal of High Energy Physics}\ }\textbf {\bibinfo
  {volume} {2004}},\ \bibinfo {pages} {009} (\bibinfo {year}
  {2004})}\BibitemShut {NoStop}%
\bibitem [{\citenamefont {Avramidy}\ and\ \citenamefont
  {Barvinsky}(1985)}]{Avramidy1985}%
  \BibitemOpen
  \bibfield  {author} {\bibinfo {author} {\bibfnamefont {I.}~\bibnamefont
  {Avramidy}}\ and\ \bibinfo {author} {\bibfnamefont {A.}~\bibnamefont
  {Barvinsky}},\ }\bibfield  {title} {\bibinfo {title} {Asymptotic freedom in
  higher-derivative quantum gravity},\ }\href
  {https://doi.org/https://doi.org/10.1016/0370-2693(85)90248-5} {\bibfield
  {journal} {\bibinfo  {journal} {Physics Letters B}\ }\textbf {\bibinfo
  {volume} {159}},\ \bibinfo {pages} {269} (\bibinfo {year}
  {1985})}\BibitemShut {NoStop}%
\bibitem [{\citenamefont {Mannheim}(2012)}]{Mannheim2012}%
  \BibitemOpen
  \bibfield  {author} {\bibinfo {author} {\bibfnamefont {P.~D.}\ \bibnamefont
  {Mannheim}},\ }\bibfield  {title} {\bibinfo {title} {Making the case for
  conformal gravity},\ }\href {https://doi.org/10.1007/s10701-011-9608-6}
  {\bibfield  {journal} {\bibinfo  {journal} {Foundations of Physics}\ }\textbf
  {\bibinfo {volume} {42}},\ \bibinfo {pages} {388} (\bibinfo {year}
  {2012})}\BibitemShut {NoStop}%
\bibitem [{\citenamefont {Yang}\ \emph {et~al.}(2013)\citenamefont {Yang},
  \citenamefont {Chen}, \citenamefont {Zhao}, \citenamefont {Li},\ and\
  \citenamefont {Liu}}]{Yang2013}%
  \BibitemOpen
  \bibfield  {author} {\bibinfo {author} {\bibfnamefont {R.}~\bibnamefont
  {Yang}}, \bibinfo {author} {\bibfnamefont {B.}~\bibnamefont {Chen}}, \bibinfo
  {author} {\bibfnamefont {H.}~\bibnamefont {Zhao}}, \bibinfo {author}
  {\bibfnamefont {J.}~\bibnamefont {Li}},\ and\ \bibinfo {author}
  {\bibfnamefont {Y.}~\bibnamefont {Liu}},\ }\bibfield  {title} {\bibinfo
  {title} {Test of conformal gravity with astrophysical observations},\ }\href
  {https://doi.org/https://doi.org/10.1016/j.physletb.2013.10.035} {\bibfield
  {journal} {\bibinfo  {journal} {Physics Letters B}\ }\textbf {\bibinfo
  {volume} {727}},\ \bibinfo {pages} {43} (\bibinfo {year} {2013})}\BibitemShut
  {NoStop}%
\bibitem [{\citenamefont {Oda}(2018)}]{Oda2018}%
  \BibitemOpen
  \bibfield  {author} {\bibinfo {author} {\bibfnamefont {I.}~\bibnamefont
  {Oda}},\ }\bibfield  {title} {\bibinfo {title} {Planck and electroweak scales
  emerging from conformal gravity},\ }\href
  {https://doi.org/10.1140/epjc/s10052-018-6289-8} {\bibfield  {journal}
  {\bibinfo  {journal} {The European Physical Journal C}\ }\textbf {\bibinfo
  {volume} {78}},\ \bibinfo {pages} {798} (\bibinfo {year} {2018})}\BibitemShut
  {NoStop}%
\bibitem [{\citenamefont {Zee}(1983)}]{Zee1983}%
  \BibitemOpen
  \bibfield  {author} {\bibinfo {author} {\bibfnamefont {A.}~\bibnamefont
  {Zee}},\ }\bibfield  {title} {\bibinfo {title} {Einstein gravity emerging
  from quantum {Weyl} gravity},\ }\href
  {https://doi.org/https://doi.org/10.1016/0003-4916(83)90286-5} {\bibfield
  {journal} {\bibinfo  {journal} {Annals of Physics}\ }\textbf {\bibinfo
  {volume} {151}},\ \bibinfo {pages} {431} (\bibinfo {year}
  {1983})}\BibitemShut {NoStop}%
\bibitem [{\citenamefont {Maldacena}(2011)}]{Maldacena2011}%
  \BibitemOpen
  \bibfield  {author} {\bibinfo {author} {\bibfnamefont {J.}~\bibnamefont
  {Maldacena}},\ }\href {https://doi.org/10.48550/ARXIV.1105.5632} {\bibinfo
  {title} {Einstein gravity from conformal gravity, {arXiv}:1105.5632}}
  (\bibinfo {year} {2011})\BibitemShut {NoStop}%
\bibitem [{\citenamefont {Anastasiou}\ and\ \citenamefont
  {Olea}(2016)}]{Anastasiou2016}%
  \BibitemOpen
  \bibfield  {author} {\bibinfo {author} {\bibfnamefont {G.}~\bibnamefont
  {Anastasiou}}\ and\ \bibinfo {author} {\bibfnamefont {R.}~\bibnamefont
  {Olea}},\ }\bibfield  {title} {\bibinfo {title} {{From conformal to Einstein
  gravity}},\ }\href {https://doi.org/10.1103/PhysRevD.94.086008} {\bibfield
  {journal} {\bibinfo  {journal} {Phys. Rev. D}\ }\textbf {\bibinfo {volume}
  {94}},\ \bibinfo {pages} {086008} (\bibinfo {year} {2016})}\BibitemShut
  {NoStop}%
\bibitem [{\citenamefont {Anastasiou}\ \emph {et~al.}(2021)\citenamefont
  {Anastasiou}, \citenamefont {Araya},\ and\ \citenamefont
  {Olea}}]{Anastasiou2021}%
  \BibitemOpen
  \bibfield  {author} {\bibinfo {author} {\bibfnamefont {G.}~\bibnamefont
  {Anastasiou}}, \bibinfo {author} {\bibfnamefont {I.~J.}\ \bibnamefont
  {Araya}},\ and\ \bibinfo {author} {\bibfnamefont {R.}~\bibnamefont {Olea}},\
  }\bibfield  {title} {\bibinfo {title} {{Einstein gravity from Conformal
  Gravity in 6D}},\ }\href {https://doi.org/10.1007/JHEP01(2021)134} {\bibfield
   {journal} {\bibinfo  {journal} {Journal of High Energy Physics}\ }\textbf
  {\bibinfo {volume} {2021}},\ \bibinfo {pages} {134} (\bibinfo {year}
  {2021})}\BibitemShut {NoStop}%
\bibitem [{\citenamefont {Jizba}\ \emph {et~al.}(2015)\citenamefont {Jizba},
  \citenamefont {Kleinert},\ and\ \citenamefont {Scardigli}}]{Jizba2015}%
  \BibitemOpen
  \bibfield  {author} {\bibinfo {author} {\bibfnamefont {P.}~\bibnamefont
  {Jizba}}, \bibinfo {author} {\bibfnamefont {H.}~\bibnamefont {Kleinert}},\
  and\ \bibinfo {author} {\bibfnamefont {F.}~\bibnamefont {Scardigli}},\
  }\bibfield  {title} {\bibinfo {title} {Inflationary cosmology from quantum
  conformal gravity},\ }\href {https://doi.org/10.1140/epjc/s10052-015-3441-6}
  {\bibfield  {journal} {\bibinfo  {journal} {Eur. Phys. J. C}\ }\textbf
  {\bibinfo {volume} {75}},\ \bibinfo {pages} {245} (\bibinfo {year}
  {2015})}\BibitemShut {NoStop}%
\bibitem [{\citenamefont {Sakharov}(1991)}]{Sakharov1991}%
  \BibitemOpen
  \bibfield  {author} {\bibinfo {author} {\bibfnamefont {A.~D.}\ \bibnamefont
  {Sakharov}},\ }\bibfield  {title} {\bibinfo {title} {Vacuum quantum
  fluctuations in curved space and the theory of gravitation},\ }\href
  {https://doi.org/10.1070/PU1991v034n05ABEH002498} {\bibfield  {journal}
  {\bibinfo  {journal} {Soviet Physics Uspekhi}\ }\textbf {\bibinfo {volume}
  {34}},\ \bibinfo {pages} {394} (\bibinfo {year} {1991})}\BibitemShut
  {NoStop}%
\bibitem [{\citenamefont {Sakharov}(2000)}]{Sakharov2000}%
  \BibitemOpen
  \bibfield  {author} {\bibinfo {author} {\bibfnamefont {A.~D.}\ \bibnamefont
  {Sakharov}},\ }\bibfield  {title} {\bibinfo {title} {Vacuum quantum
  fluctuations in curved space and the theory of gravitation},\ }\href
  {https://doi.org/10.1023/A:1001947813563} {\bibfield  {journal} {\bibinfo
  {journal} {General Relativity and Gravitation}\ }\textbf {\bibinfo {volume}
  {32}},\ \bibinfo {pages} {365} (\bibinfo {year} {2000})}\BibitemShut
  {NoStop}%
\bibitem [{\citenamefont {Adler}(1982)}]{Adler1982}%
  \BibitemOpen
  \bibfield  {author} {\bibinfo {author} {\bibfnamefont {S.~L.}\ \bibnamefont
  {Adler}},\ }\bibfield  {title} {\bibinfo {title} {Einstein gravity as a
  symmetry-breaking effect in quantum field theory},\ }\href
  {https://doi.org/10.1103/RevModPhys.54.729} {\bibfield  {journal} {\bibinfo
  {journal} {Rev. Mod. Phys.}\ }\textbf {\bibinfo {volume} {54}},\ \bibinfo
  {pages} {729} (\bibinfo {year} {1982})}\BibitemShut {NoStop}%
\bibitem [{\citenamefont {Visser}(2002)}]{Visser2002}%
  \BibitemOpen
  \bibfield  {author} {\bibinfo {author} {\bibfnamefont {M.}~\bibnamefont
  {Visser}},\ }\bibfield  {title} {\bibinfo {title} {Sakharov's induced
  gravity: A modern perspective},\ }\href
  {https://doi.org/10.1142/S0217732302006886} {\bibfield  {journal} {\bibinfo
  {journal} {Modern Physics Letters A}\ }\textbf {\bibinfo {volume} {17}},\
  \bibinfo {pages} {977} (\bibinfo {year} {2002})}\BibitemShut {NoStop}%
\bibitem [{\citenamefont {Altshuler}(2021)}]{Altshuler2021}%
  \BibitemOpen
  \bibfield  {author} {\bibinfo {author} {\bibfnamefont {B.~L.}\ \bibnamefont
  {Altshuler}},\ }\bibfield  {title} {\bibinfo {title} {{Andrei Sakharov's
  research work and modern physics}},\ }\href
  {https://doi.org/10.3367/UFNe.2021.02.038946} {\bibfield  {journal} {\bibinfo
   {journal} {Physics-Uspekhi}\ }\textbf {\bibinfo {volume} {64}},\ \bibinfo
  {pages} {427} (\bibinfo {year} {2021})}\BibitemShut {NoStop}%
\bibitem [{\citenamefont {Konopka}\ \emph {et~al.}(2006)\citenamefont
  {Konopka}, \citenamefont {Markopoulou},\ and\ \citenamefont
  {Smolin}}]{Konopka2006}%
  \BibitemOpen
  \bibfield  {author} {\bibinfo {author} {\bibfnamefont {T.}~\bibnamefont
  {Konopka}}, \bibinfo {author} {\bibfnamefont {F.}~\bibnamefont
  {Markopoulou}},\ and\ \bibinfo {author} {\bibfnamefont {L.}~\bibnamefont
  {Smolin}},\ }\href {https://doi.org/10.48550/ARXIV.HEP-TH/0611197} {\bibinfo
  {title} {Quantum graphity, {arXiv}:hep-th/0611197}} (\bibinfo {year}
  {2006})\BibitemShut {NoStop}%
\bibitem [{\citenamefont {Verlinde}(2011)}]{Verlinde2011}%
  \BibitemOpen
  \bibfield  {author} {\bibinfo {author} {\bibfnamefont {E.}~\bibnamefont
  {Verlinde}},\ }\bibfield  {title} {\bibinfo {title} {On the origin of gravity
  and the laws of {N}ewton},\ }\href {https://doi.org/10.1007/JHEP04(2011)029}
  {\bibfield  {journal} {\bibinfo  {journal} {Journal of High Energy Physics}\
  }\textbf {\bibinfo {volume} {2011}},\ \bibinfo {pages} {29} (\bibinfo {year}
  {2011})}\BibitemShut {NoStop}%
\bibitem [{\citenamefont {Sindoni}(2012)}]{Sindoni2012}%
  \BibitemOpen
  \bibfield  {author} {\bibinfo {author} {\bibfnamefont {L.}~\bibnamefont
  {Sindoni}},\ }\bibfield  {title} {\bibinfo {title} {Symmetry, integrability
  and geometry: Methods and applications},\ }\href
  {https://doi.org/http://dx.doi.org/10.3842/SIGMA.2012.027} {\bibfield
  {journal} {\bibinfo  {journal} {SIGMA}\ }\textbf {\bibinfo {volume} {8}},\
  \bibinfo {pages} {027} (\bibinfo {year} {2012})}\BibitemShut {NoStop}%
\bibitem [{\citenamefont {Berman}\ \emph {et~al.}(2022)\citenamefont {Berman},
  \citenamefont {Collins},\ and\ \citenamefont {Persson}}]{Berman2022}%
  \BibitemOpen
  \bibfield  {author} {\bibinfo {author} {\bibfnamefont {R.~J.}\ \bibnamefont
  {Berman}}, \bibinfo {author} {\bibfnamefont {T.~C.}\ \bibnamefont
  {Collins}},\ and\ \bibinfo {author} {\bibfnamefont {D.}~\bibnamefont
  {Persson}},\ }\bibfield  {title} {\bibinfo {title} {{Emergent Sasaki-Einstein
  geometry and AdS/CFT}},\ }\href {https://doi.org/10.1038/s41467-021-27951-9}
  {\bibfield  {journal} {\bibinfo  {journal} {Nature Communications}\ }\textbf
  {\bibinfo {volume} {13}},\ \bibinfo {pages} {365} (\bibinfo {year}
  {2022})}\BibitemShut {NoStop}%
\bibitem [{\citenamefont {Sakharov}(1975)}]{Sakharov1975}%
  \BibitemOpen
  \bibfield  {author} {\bibinfo {author} {\bibfnamefont {A.~D.}\ \bibnamefont
  {Sakharov}},\ }\bibfield  {title} {\bibinfo {title} {Spectral density of
  eigenvalues of the wave equation and vacuum polarization},\ }\href
  {https://doi.org/10.1007/BF01036152} {\bibfield  {journal} {\bibinfo
  {journal} {Theoretical and Mathematical Physics}\ }\textbf {\bibinfo {volume}
  {23}},\ \bibinfo {pages} {435} (\bibinfo {year} {1975})}\BibitemShut
  {NoStop}%
\bibitem [{\citenamefont {Christensen}\ and\ \citenamefont
  {Duff}(1980)}]{Christensen1980}%
  \BibitemOpen
  \bibfield  {author} {\bibinfo {author} {\bibfnamefont {S.}~\bibnamefont
  {Christensen}}\ and\ \bibinfo {author} {\bibfnamefont {M.}~\bibnamefont
  {Duff}},\ }\bibfield  {title} {\bibinfo {title} {Quantizing gravity with a
  cosmological constant},\ }\href
  {https://doi.org/https://doi.org/10.1016/0550-3213(80)90423-X} {\bibfield
  {journal} {\bibinfo  {journal} {Nuclear Physics B}\ }\textbf {\bibinfo
  {volume} {170}},\ \bibinfo {pages} {480} (\bibinfo {year}
  {1980})}\BibitemShut {NoStop}%
\bibitem [{\citenamefont {Birrell}\ and\ \citenamefont
  {Davies}(1982)}]{Birrell_Davies_book}%
  \BibitemOpen
  \bibfield  {author} {\bibinfo {author} {\bibfnamefont {N.~D.}\ \bibnamefont
  {Birrell}}\ and\ \bibinfo {author} {\bibfnamefont {P.~C.~W.}\ \bibnamefont
  {Davies}},\ }\href {https://doi.org/10.1017/CBO9780511622632} {\emph
  {\bibinfo {title} {Quantum Fields in Curved Space}}},\ Cambridge Monographs
  on Mathematical Physics\ (\bibinfo  {publisher} {Cambridge University
  Press},\ \bibinfo {year} {1982})\BibitemShut {NoStop}%
\bibitem [{\citenamefont {Kehagias}\ \emph {et~al.}(2021)\citenamefont
  {Kehagias}, \citenamefont {Partouche},\ and\ \citenamefont {{de
  Vaulchier}}}]{Kehagias2021}%
  \BibitemOpen
  \bibfield  {author} {\bibinfo {author} {\bibfnamefont {A.}~\bibnamefont
  {Kehagias}}, \bibinfo {author} {\bibfnamefont {H.}~\bibnamefont
  {Partouche}},\ and\ \bibinfo {author} {\bibfnamefont {B.}~\bibnamefont {{de
  Vaulchier}}},\ }\bibfield  {title} {\bibinfo {title} {Induced {Einstein}
  gravity from infinite towers of states},\ }\href
  {https://doi.org/https://doi.org/10.1016/j.nuclphysb.2021.115312} {\bibfield
  {journal} {\bibinfo  {journal} {Nuclear Physics B}\ }\textbf {\bibinfo
  {volume} {964}},\ \bibinfo {pages} {115312} (\bibinfo {year}
  {2021})}\BibitemShut {NoStop}%
\bibitem [{\citenamefont {Brewin}(2009)}]{Brewin2009}%
  \BibitemOpen
  \bibfield  {author} {\bibinfo {author} {\bibfnamefont {L.}~\bibnamefont
  {Brewin}},\ }\bibfield  {title} {\bibinfo {title} {Riemann normal coordinate
  expansions using {C}adabra},\ }\href
  {https://doi.org/10.1088/0264-9381/26/17/175017} {\bibfield  {journal}
  {\bibinfo  {journal} {Classical and Quantum Gravity}\ }\textbf {\bibinfo
  {volume} {26}},\ \bibinfo {pages} {175017} (\bibinfo {year}
  {2009})}\BibitemShut {NoStop}%
\bibitem [{\citenamefont {Teyssandier}\ \emph {et~al.}(2008)\citenamefont
  {Teyssandier}, \citenamefont {Poncin-Lafitte},\ and\ \citenamefont
  {Linet}}]{Teyssandier2008}%
  \BibitemOpen
  \bibfield  {author} {\bibinfo {author} {\bibfnamefont {P.}~\bibnamefont
  {Teyssandier}}, \bibinfo {author} {\bibfnamefont {C.~L.}\ \bibnamefont
  {Poncin-Lafitte}},\ and\ \bibinfo {author} {\bibfnamefont {B.}~\bibnamefont
  {Linet}},\ }\bibinfo {title} {A universal tool for determining the time delay
  and the frequency shift of light: Synge's world function},\ in\ \href
  {https://doi.org/10.1007/978-3-540-34377-6_6} {\emph {\bibinfo {booktitle}
  {Lasers, Clocks and Drag-Free Control: Exploration of Relativistic Gravity in
  Space}}},\ \bibinfo {editor} {edited by\ \bibinfo {editor} {\bibfnamefont
  {H.}~\bibnamefont {Dittus}}, \bibinfo {editor} {\bibfnamefont
  {C.}~\bibnamefont {L{\"a}mmerzahl}},\ and\ \bibinfo {editor} {\bibfnamefont
  {S.~G.}\ \bibnamefont {Turyshev}}}\ (\bibinfo  {publisher} {Springer Berlin
  Heidelberg},\ \bibinfo {address} {Berlin, Heidelberg},\ \bibinfo {year}
  {2008})\ pp.\ \bibinfo {pages} {153--180}\BibitemShut {NoStop}%
\bibitem [{\citenamefont {Strichartz}(1988)}]{Strichartz1988}%
  \BibitemOpen
  \bibfield  {author} {\bibinfo {author} {\bibfnamefont {R.~S.}\ \bibnamefont
  {Strichartz}},\ }\bibfield  {title} {\bibinfo {title} {Linear algebra of
  curvature tensors and their covariant derivatives},\ }\href
  {https://doi.org/10.4153/CJM-1988-046-7} {\bibfield  {journal} {\bibinfo
  {journal} {Canadian Journal of Mathematics}\ }\textbf {\bibinfo {volume}
  {40}},\ \bibinfo {pages} {1105–1143} (\bibinfo {year} {1988})}\BibitemShut
  {NoStop}%
\end{thebibliography}%

\end{document}